\documentclass[
superscriptaddress,
amsmath,amssymb,
preprint,
 preprintnumbers,
onecolumn,
pra,
aps,
floatfix,longbibliography
]{revtex4-2}

\usepackage{graphicx}
\usepackage{dcolumn}
\usepackage{bm}
\usepackage{chngcntr} 
\usepackage[pdftex,colorlinks=true, linkcolor = blue, citecolor=blue,urlcolor=blue, bookmarksnumbered=true, bookmarksopen=true]{hyperref}
\usepackage[mathlines]{lineno}
\usepackage{physics}
\usepackage{tikz}
\usetikzlibrary{quantikz}
\usepackage{algpseudocode}
\usepackage{algorithm}
\usepackage{lineno}
\usepackage{booktabs}
\usepackage{adjustbox}
\usepackage{soul}
\usepackage{subcaption}
\usepackage{multirow}
\usepackage{caption}
\begin{document}

\title{Quantum Algorithms for State Preparation and Data Classification based on Stabilizer Codes}

\author{Pejman Jouzdani}
\email{Corresponding author, email: jouzdanip@fusion.gat.com}
\affiliation{General Atomics, 3550 General Atomics Ct, San Diego, CA 92121, USA}
\author{H. Arslan Hashim}
\affiliation{Department of Physics, University of Central Florida, Orlando, FL 32816, USA}
\author{Eduardo R. Mucciolo}
\email{Corresponding author, email: Eduardo.Mucciolo@ucf.edu}
\affiliation{Department of Physics, University of Central Florida, Orlando, FL 32816, USA}

\date{\today}

\begin{abstract}
Quantum error correction (QEC) is a way to protect quantum information against noise. It consists of encoding input information into entangled quantum states known as the code space. Furthermore, to classify if the encoded information is corrupted or intact, a step known as syndrome detection is performed. For stabilizer codes, this step consists of measuring a set of stabilizer operators. In this paper, inspired by the QEC approach, and specifically stabilizer codes, we propose a prototype quantum circuit model for classification of classical data. 
The core quantum circuit can be considered as a \emph{quantum perceptron} where the classification is based on syndrome detection. 
In this proposal, a quantum perceptron is realized by one stabilizer as part of a stabilizer code, while a quantum neural network (QNN) layer is realized by a stabilizer code which consists of many stabilizers. The concatenation of stabilizer codes results in complex QNNs. The QNN is trained by performing measurements and optimization of a set of parameterized stabilizers. We demonstrate the concept numerically.
In this paper we also consider the first challenge to most applications of quantum computers, including data classification, which is to load data into the memory of the quantum computer. This loading amounts 
to representing the data as a quantum state, i.e., quantum state preparation. 
An exact amplitude encoding algorithm requires a circuit of exponential depth. We introduce an alternative recursive algorithm which approximates amplitude encoding with only a polynomial number of elementary gates. We name it recursive approximate-scheme algorithm (RASA).
\end{abstract}



\maketitle

\section{Introduction}
\label{sec:introduction}

In the emerging era of quantum computers, quantum algorithms offer solutions that exploit the power of quantum entanglement. In recent years, a 
large body of research has explored the problem of quantum machine learning (QML) \cite{Mitarai2018, Schuld2019}, 
where the goal is to perform machine learning tasks such as classifications on quantum computers \cite{Lloyd2020, Bae2015}, potentially achieving the so-called quantum advantage \cite{Daley2022}.

An ongoing effort in QML is the pursuit of the quantum equivalent of artificial neural networks. Attempts to design quantum neural networks  
have been reported \cite{Schuld2014A, cao2017quantum, Wan2017, Rebentrost2018}.
The essential building block of an artificial neural network \cite{rumelhart1986learning, Lecun1998, Krizhevsky2012} is the perceptron 
\cite{Rosenblatt_1957_6098}.
A perceptron is a statistical device where a set of 
inputs $\vec{x}=(x_1, \cdots, x_t)$ is entered and a binary output  $y\in\{-1, +1\}$ is produced.
Naturally, the pursuit of a quantum version of a perceptron, i.e., a quantum perceptron (QP), has led to several approaches. In Ref. \cite{Schuld2015},  the authors use a quantum phase estimation (QPA) algorithm to label the input data, with the output being a function of the realized phase. 
In a recent work \cite{Tacchino2019}, a quantum circuit model was designed to replicate the function of a classical perceptron for a special case where the parameters and input data take the discrete values $\pm 1$. 

QEC was developed to combat noise and safeguard quantum information. 
The idea is that the quantum information content of some logical qubits can be \emph{encoded} in the Hilbert space of a larger number of physical qubits. 
When errors affect the state of these physical qubits, the original quantum information content of the logical qubits can be recovered with some probability. In the stabilizer formalism, a quantum code space $\ket{\psi}$ is defined by the quantum states (living in the space of physical qubits) stabilized by a set of operators $\{G_i\}$, i.e., $G_i \ket{\psi} = (+1)\ket{\psi}$ for all $i$. When quantum information is stored in the code space and a correctable error occurs, measurements of some of the stabilizers will return the value $-1$ (the \emph{syndrome}), indicating the type of error that has corrupted the information. The error can then be removed by applying a recovery operation to the quantum state of the physical qubits. 

Interestingly, the way quantum states are labeled based on the detected error in the stabilizer formalism is similar to the concept of a classical perceptron: When classical information $\vec{x}$ is received in the form of a quantum state $\ket{x}$, we can imagine that $\ket{x}$ 
is the result of some quantum encoding procedure unknown to us. 
We can further consider that the encoding is specified by a set of unknown stabilizers. 
By using a training data set, suppose that the stabilizer set originally used to encode the information is approximately identified. 
Then, any given arbitrary input information can be labeled by syndrome measurements, i.e., by interpreting the syndrome as some output (or label) $y=\pm 1$, similar to a classical perceptron.

Such a viewpoint, to our best knowledge, has not been explored in the context of QNN. Some previous literature touched on similar concepts but did not propose QEC as basis for QNN. Reference \cite{Manin2018} suggested error correction as the fundamental neural network activity in the brain. Recently, Ref. \cite{Anschuetz2023} used the concept of stabilizer measurements to analyze the expressive power of QNNs.

The challenge in our approach is to approximately learn the unknown encoding protocol, or equivalently, to identify the stabilizer set. Consider a supervised learning scenario: We are given a number of sample data (a training set) that are prepared as quantum states  $\{ \ket{x^{(1)}}, \cdots \}$ 
and have associated labels $\{y^{(1)}, \cdots \}$.
Our approach to find an approximate set of stabilizers is 
to consider parameterized stabilizers where the error between the predicted labels and actual labels $y$ can be used to adjust the parameters. 
The details will be discussed in the following sections.

In the context of QEC, the number of errors that can be corrected, i.e., the code distance, is well understood and quantifiable. When conceptualizing a group of quantum perceptrons as a stabilizer code, code distance is interpreted as the ability to differentiate multiple classes of data. Similarly, a QNN can be understood as the analog to the quantum code concatenation used in fault-tolerant quantum computation. This correspondence merits further investigation and the present work explores the connection between QNN and QEC.

For practical reasons, the paper also focuses on quantum state preparation. Quantum state preparation is the main step in most quantum computing applications, and is considered a limiting factor in quantum algorithm performance. Applications in machine learning such as data classification always begin with the preparation of a quantum data set. It is known that quantum computers cannot \emph{import} data from external sources; the user must specify a way to prepare the data as a quantum state \cite{broughton2021tensorflow}. As an example, we mention quantum algorithms designed to explore application of quantum computing in finance \cite{herman2022survey, Miyamoto2022}. 

Attempts for optimal state preparation have been suggested in the recent literature. Reference \cite{Zhang2022PhysRevLett129230504} reduces circuit depth by deploying overhead ancillary qubits. Alternatively, when the input is drawn from a \emph{continuous function}, an algorithm with constant query complexity is proposed in Ref. \cite{rattew2022preparing}. Another example is the proposal in Ref. \cite{marinsanchez2021quantum} based on the Grover-Rudolph algorithm \cite{grover2002creating}. Yet, in practice, inputs of a classification problem are not necessary drawn from an analytical function, limiting the applicability of some of these proposals. For example, in the classification of hand-written numbers the input is an array of image pixels.

The physics community has put forward proposals for quantum state preparation based on a matrix-product-state (MPS) representation \cite{Ran2020PhysRevA101032310, Dilip2022PhysRevResearch}. There are perhaps two main practical challenges in these proposals. First there is the time complexity to represent matrices in terms of one-qubit and two-qubit gates \cite{Melnikov_2023, kukliansky2023qfactor}, although optimization is likely possible. Second and more worrisome is the number of elementary gates required to implement a matrix. As the auxiliary bond dimensions in the MPS increase (for better resolution of the approximated input data), the number of elementary gates increases exponentially, resulting in impractically deep circuits.

Other proposals also exist. Some are based on probabilistic quantum memories \cite{Trugenberger2000, Giovannetti2008PhysRevLett100160501}. Another approach, designed for Hamiltonian simulations, is based on a matrix analysis and achieves polynomial complexity in quantum circuit reduction \cite{Camps_2022}. Another related work is Ref. \cite{Peng2022PhysRevA106012412}. 

In this paper, we present a recursive amplitude approximation, namely, a unitary evolution that loads the (normalized) classical information $\vec{x}=(x_1, \cdots, x_t)$ into qubits. 
While the physical resource for the encoding,  the number of qubits, stays $\mathcal{O}(\log_2{(t)})$, the depth of the circuit is $\mathcal{O}(poly(\log_2{(t)})$. 
This approximation scheme uses the contrast between different parts of the data $\vec{x}$ to shrink the exact amplitude encoding algorithm \cite{Shende2006}.

The rest of paper is organized as follows. In Sec. \ref{sec-QP}, after a quick review of stabilizer codes, a quantum perceptron out of a parity check and the construction of a QNN for classification are introduced and explained in details. 
In Sec. \ref{sec-results}, numerical simulations are provided where hand-written images of digits are classified after training the model. In Sec. \ref{sec-QSP} we provide a brief introduction to the amplitude encoding algorithm often used in literature. In Sec \ref{sec-QSP-approx}, we introduce the recursive approximate-scheme algorithm (RASA). The algorithm is discussed and formally presented. In this section, we also made clear the foundational connection of our recursive approach and our previous work \cite{Jouzdani2022}. The paper concludes in Sec. \ref{sec-summary} with a summary and discussion of future directions. 
Appendix \ref{sec:appendix1} reviews details of the standard amplitude encoding algorithm. Appendix \ref{sec:appendix2}  contains the subroutines that are used in the main text as part of RASA, and 
Appendix \ref{sec:appendix3} provides a step-by-step example where steps in the recursion approach are mapped out.

\section{Building a Predictive model from a stabilizer code}
\label{sec-QP}

In this section, for the reader's benefit, we provide a short introduction to stabilizer codes in the context of quantum error correction before describing how they can be adopted for a predictive model. We then introduce our model of a quantum perceptron.

\subsection{Stabilizer Codes and Data Classification}
\label{sec-QEC}

Stabilizer codes \cite{gottesman1997stabilizer} encompass most known quantum error correction codes and are quantum generalizations of classical binary linear codes \cite{nielsen_chuang_2010}. A stabilizer code encodes $k$ logical 
qubits in $n$ physical qubits. The code space $\mathcal{L}$
is defined as a subspace of the Hilbert space of $n$ physical qubits $\mathcal{H}^{\otimes n}$ which is invariant under the action of the stabilizer group. Denoting the stabilizer group by $\mathcal{S}$,
\begin{eqnarray}
    \mathcal{L} =    
    \{
    \ket{\phi} 
    \in 
    \mathcal{H}^{\otimes n}:
    \, 
    P \, \ket{\phi} =\ket{\phi},
    \,
    \forall P 
    \in 
    \mathcal{S}
    \}
\end{eqnarray}
The stabilizer group $\mathcal{S}$ is  generated by a finite set of $m$ commuting generators, $\{G_1,\cdots, G_m\}$ ($m<n$). 
Formally, $\mathcal{S}$ is considered a subgroup of the
Pauli group, and thus a generator $G$ is a tensor product of Pauli operators multiplied by a phase that can be $\pm 1$ or $\pm i$, with $-I$ excluded \cite{Terhal2015}. 

If there are $m$ linearly independent generators in $\mathcal{S}$, then $k=n-m$, 
and the code space $\mathcal{L}$ has dimension $2^k$.  Generators allow one to perform parity check on the encoded information and detect a \emph{correctable error}. For any stabilizer code there exists a separate set of operators known as logical operators that commute with elements in $\mathcal{S}$ but are not generated by the generators of $\mathcal{S}$. The logical operators allow one to manipulate the encoded information. They also establish the distance of the code, i.e., the number of errors that can be detected and corrected by the stabilizers. 

Now consider classical input data prepared  as a quantum state $\ket{\psi} \in \mathcal{H}^{\otimes n}$. Our approach to classify this input relies on hypothesizing that $\ket{\psi}$ was originally encoded according to a stabilizer code such that the original encoder can be characterized by a set of generators.
We also make the assumption that encoded states might have been altered and corrupted before they arrived. 
Our goal is to approximate the generator set such that a classification can be performed at low computational cost. If we knew that the quantum encoder had a set of generators $\{G_1, \cdots, G_m\}$, by measuring them via parity check we could straightforwardly classify the input quantum state, i.e., specify the type of error that had "corrupted it". This approach works up to the distance of the assumed underlying stabilizer code. However, we do not know the presumed generator set and thus must find it.

Finding the exact generators is a difficult task as the Pauli group on $n$ qubits has an exponential number of elements. To restrict the search, we first assume a stabilizer code generated by a limited set, hereafter denoted by $\mathcal{S}=<G_1, \cdots, G_m>$. 
Notice that for any arbitrary unitary transformation $U$ defined on  $\mathcal{H}^{\otimes n}$, the linear independence of the generators is preserved under the transformation $U$. Thus, an entire class of stabilizer codes can be generated by defining
$\mathcal{S}(\theta) = U^{\dagger}(\theta) \, \mathcal{S} \, U(\theta) = < 
U^{\dagger}(\theta)
G_1
U(\theta),
\cdots,
U^{\dagger}(\theta)
G_m
U(\theta)
>,
$
where $U(\theta)$ can be thought of as a product of a limited number of parametric quantum gates, with the vector of parameters given by $\theta$.  

In addition, to find the original stabilizer code we must have access to a set of training data that consists of input quantum states $\ket{\phi}$ and their corresponding labels. The label in our interpretation identifies the type of error that has occurred. By adjusting the parameters in $U(\theta)$ we can minimize the error between the parity check measurements and the actual labels over the training set. Once the optimal transformation $U(\theta$) is found, or, equivalently, the encoding protocol $\mathcal{S}(\theta)$, it can be used to make predictions on unlabeled quantum states; the result is a predictive model.

\subsection{Quantum Perceptron}
\label{sec-QP-QEC}
To illustrate the relation between the stabilizer code hypothesis and perceptrons, consider a situation where a set of inputs $\vec{x}$ and an associated label $y\in \{-1,1\}$ are provided. Assume without loss of generality that the 
inputs $\vec{x}$ are prepared as a normalized quantum state $\ket{x}$.
Assume that $\ket{x}$ was originally encoded by an unknown stabilizer code and that 
this stabilizer code is defined by a single stabilizer operator $G$. Initially, since $\ket{x}$ is assumed to be prepared in the code space, $G\ket{x} = +1\,\ket{x} $. If the information in the quantum state is altered, then a 
measurement of $G$ will only yield $+1$ with some probability, and will yield $-1$ otherwise. 
This behavior is what is expected from the quantum equivalent of a perceptron. 

In fact, there is a clear path to combine multiple of these units toward a quantum neural network (QNN). In the example above, consider a situation where the code is actually defined by two stabilizers $G^{(h)}_1$ and $G^{(h)}_2$, and not just one. In QEC, a measurement of the two operators is equivalent to investigating the occurrence of two types of errors (a bit-flip and a phase-flip, for instance). 
Recall that the input data points have only one label $y$. 
Thus we must, by some means, infer one label out of the measurement of two stabilizers $G^{(h)}_1$ and $G^{(h)}_2$. 
A way around this is to consider the output measurements 
of $G^{(h)}_1$ and $G^{(h)}_2$ as a set of \emph{encoded} inputs to a final quantum perceptron defined by a separate stabilizer.
The latter code is  defined by a stabilizer that is  
different from $G^{(h)}_1$ and $G^{(h)}_2$; let us denote it by 
$G^{(o)}$. 
Then, the measurement of $G^{(o)}$ yields the final prediction of the  label of the input data.
In practice the stabilizers are unknown and thus a parameterized set $\{G(\theta)\}$ should be considered. The parameters are tuned using the training data set. 
The training is done by having access to a set of data with known labels (supervised learning). For the purpose of training of the perceptron, the 
prediction is compared with the exact label $y$ of the training data. 
The difference is then used to build a cost function and minimization is done to find the best values of the circuit parameters, a common practice in variational quantum algorithms. 

\begin{figure*}
\includegraphics[width=\textwidth]{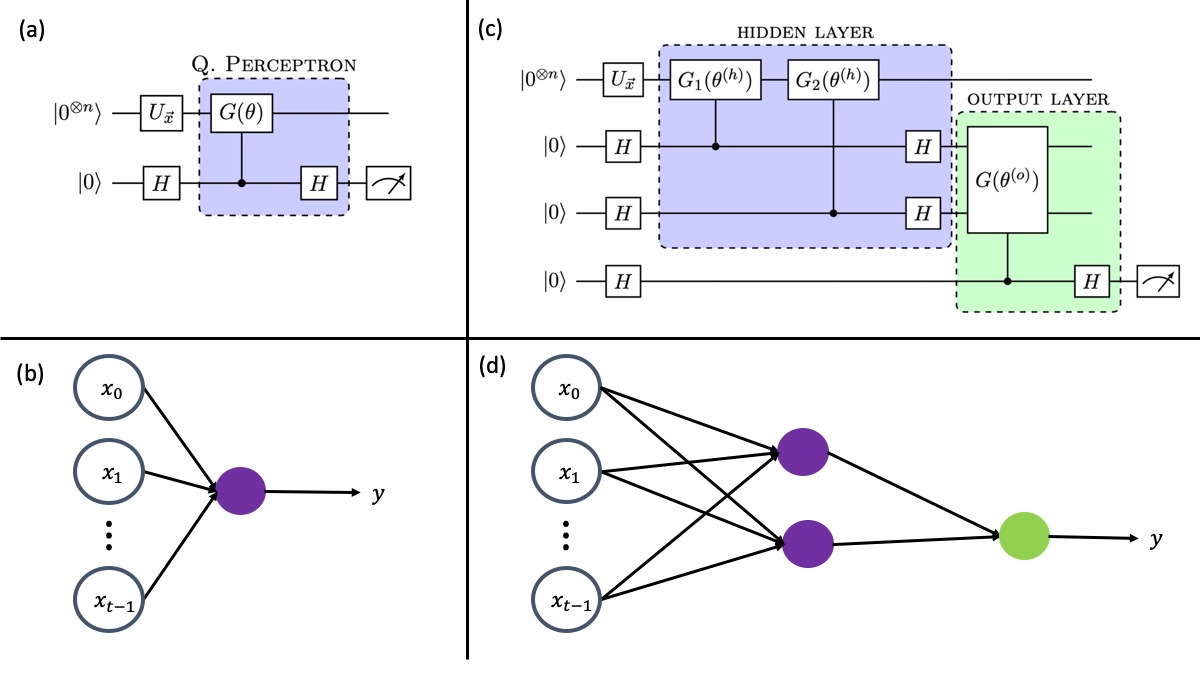}
\caption{
(a) In our proposed QNN 
a single quantum perceptron is defined by a single stabilizer 
$G(\theta)= 
U^{\dagger}(\theta)\, G \, U(\theta)$. In this comparison a quantum perceptron classifies an input data as corrupted/uncorrupted based on the error that $G(\theta)$ corresponds to. 
(b) The classical perceptron equivalent to the quantum circuit in (a). 
(c) A \emph{hidden layer} 
is a collection of QPs that can be measured simultaneously, that is a stabilizer code. In this example $\mathcal{S}^{(h)}= <G_1(\theta^{(h)}), G_2(\theta^{(h)})>$ is used to perform the first round of syndrome detection. Furthermore, a multi-layer QNN is developed by concatenating different stabilizer codes.
In this example the output of $\mathcal{S}^{(h)}$ is passed to 
$\mathcal{S}^{(o)}= <G(\theta^{(o)})>$ where the final decision on the syndrome, (the final label) is made. 
(d) The equivalent classical ANN to the quantum circuit in (c).
Our proposed approach to QNNs
opens the possibility to quantify the  classification power of a hidden layer by studying the code distance of the associated stabilizer codes.}
\label{fig-4-1}
\end{figure*}

In a sense, one can interpret the first round of syndrome detection (measurement of $G^{(h)}_1$ and $G^{(h)}_2$) as a \emph{hidden layer} of a quantum neural network (QNN), while the second round of syndrome detection (measurement of $G^{(o)}$) as the output layer of the QNN. 
A summary of the above description of QP and QNN is shown in Fig. \ref{fig-4-1}.

In our approach toward QNNs, a single perceptron in a layer of QNN corresponds to measuring a single stabilizer (parity check). The first step is clearly loading of the input data into the qubits. This is done by an oracle $U_{\vec{x}}$. 
Figure \ref{fig-4-1}(a) shows the concept of a single quantum perceptron. A QP of the quantum circuit corresponds to measuring a single stabilizer $G(\theta)$ by a swap test. 
$\theta$ is a short notation for a range of parameters, i.e., $\vec{\theta}$. 
Figure \ref{fig-4-1}(b) shows the equivalent classical perceptron.
In Fig. \ref{fig-4-1}(c), two QPs are combined to comprise a hidden layer. The hidden layer is essentially a stabilizer code that in this example has two generators.
The measured parities from one layer are inputs to the the next layer. 
Figure \ref{fig-4-1}(d) shows the classical equivalent artificial neural network (ANN). The exact parametric structure of a stabilizer $G(\theta)$ in terms of elementary gates is not shown; we return to these details later in Sec.~\ref{sec-results}.

\section{Hand-written Digit Recognition}
\label{sec-results}

\subsection{The single stabilizer case}
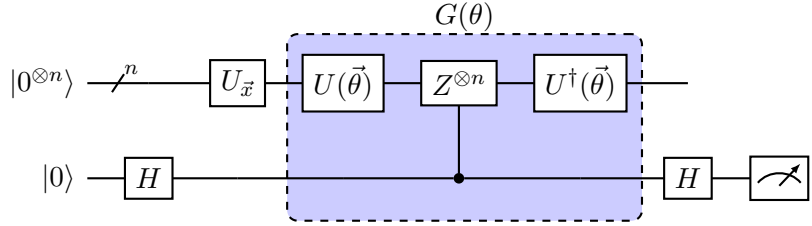
\begin{figure*}
\begin{quantikz}
\lstick{$\ket{0^{\otimes n}}$} 
& \qwbundle{n}
&\gate{U_{\vec{x}}}
& \gate{U(\vec{\theta})}
\gategroup[2,steps=3,style={dashed,
                   rounded corners,fill=blue!20, inner xsep=2pt},
                   background]{{\sc $G(\theta)$}}
& \gate{Z^{\otimes n}}
& \gate{U^\dagger (\vec{\theta})}
& \qw
\\
\lstick{$\ket{0}$} 
& \gate{H}
& \qw
& \qw
& \ctrl{-1}
& \qw
& \gate{H}
& \meter{}
\end{quantikz}
\caption{Layout of a single quantum perceptron.
Here the single stabilizer is $G(\theta) =U^\dagger(\vec{\theta}) \, \left( \prod_{q=1}^{n} Z_q \right) \, U(\vec{\theta})$.
For $\theta=0$, the syndrome corresponds to the detection of 
a bit-flip error on an odd number of qubits. 
The inclusion of the parameters $\vec{\theta}$ allows for the exploration of more generic errors. 
The measuremnt box at the far right of the last quantum wire of the circuit represents measurement of $\langle Z_{n+1}\rangle$.
}

\label{fig-5-1}
\end{figure*}

To illustrate the classification power of a single QP based on the concept of a stabilizer, here we consider the classification of hand-written numbers.
The base stabilizer  of this QP is $G = \prod_{q=1}^{q=n} Z_q$ while the parametric part is $ U(\theta) = \prod_{q=1}^{n} R_y(\theta_q)$; product of single-qubit $Y$-rotation gates. By varying the parameters 
$
\theta \equiv \{ \theta_1,\cdots,\theta_n
\}
$ 
one can explore a range of stabilizers for this QP, namely, $G(\theta) = U(\theta) G U^\dagger (\theta)$. 
The explicit implementation is shown Fig. \ref{fig-5-1}, which is a special case of the circuit shown in Fig. \ref{fig-4-1}(a).

The data to be classified are images of hand-written digits. We used a data set from Ref. \cite{OptRecogHwrittenDig1998} that contains images of ten different digits. 
Each image is $8\times 8$ pixels, with a pixel value from $0-16$. 
We reshape the image from $8\times8$ to $64\times1$ and normalize it before feeding it into the classifier. Since in our implementation a single QP is a parity check and has two outcomes, just as in a classical perceptron, we can only classify two digits with one QP. 
In this case, the data set is filtered to contain only images of two digits such as $0$ and $1$ or $8$ and $6$, etc.
Additionally, in this numerical experiment it is assumed that an oracle  $U_{\vec{x}}$ loads (i.e., encodes) an input image data into the state of $n$ qubits. 
In this study, $n=6$ and the input is an $8\times 8$ pixel image.

\emph{Data selection and training.--}
The data set with images of hand-written digits of $0$ and $1$ contained a total number of $360$ images. The data set is further ramdomly split with a 50:50 ratio into a subset of training data with $n_s = 180$ images and a testing subset of $n_v = 180$ sample images.

The quantum circuit model (Fig. \ref{fig-5-1}) was then trained on  the training subset. 
The training began by assigning some initial values to the gate parameters $\vec{\theta}$. Then, the  expectation value $\langle Z_{n+1} \rangle$ was computed for every input image, yielding a value between $0$ and $1$ that corresponds to measuring the the parity operator $(\mathbb{I}+G(\theta))/2$. The cost function was defined as the accumulation of the parity measurements with respect to all the input images in the training subset,
\begin{eqnarray}
C (\vec{\theta}) = \frac{1}{n_s}
\sum_s 
\left\vert \langle  Z_{n+1} \rangle_s  (\vec{\theta}) - y_s \right\vert^2 ,
\label{eq-cost}
\end{eqnarray}  
where $y_s\in \{0,1\}$ is the label associated with the image $s$.

\emph{Model optimization.--} 
A classical optimizer was used to minimize the cost function in Eq. (\ref{eq-cost}). 
For every input value $\vec{\theta}$, the circuit was run for all images in the training subset. This task can be done in parallel with different quantum processors. In this numerical demonstration we imitate this by executing circuits, corresponding to different input images, in parallel, which speeds up the computation significantly. 

Further, the cost function above was computed by averaging over multiple inputs and for every input $\vec{\theta}$.
The fact that the parametric circuit is  built out of only single-qubit gates, and the averaging over a training set in Eq. (\ref{eq-cost}), suggests that $C(\vec{\theta})$ has a smooth manifold and  optimization is expected avoid barren plateaux.

Due to the simplicity of the parametric structure of the circuit, the optimization was fast. During the training, we divide the training data set into several batches, and for each batch, quantum circuits are run in parallel. It took $13$ seconds on average for training on the computer with a processor Intel(R) Core(TM) i$5-9500$ CPU $@$ $3.00$ GHz, $3000$ MHz, $6$ cores, and $6$ logical processors.

In this study, General-purpose Optimization algorithms (GPO) are used. Specifically, we used the \textit{minimize} subroutine available as part of the Python scipy.optimize package for the minimization \cite{SciPyNMeth2020}. Several optimization methods available in this numerical package were tested. Among them, COBYLA, and L-BFGS-B  were more effective and resulted in the best convergence on the training subset. Here, only results for the COBYLA method are presented. The trained circuit model was then used to make predictions on the $n_v$ validation samples.

\emph{Results.-} 
Out of $180$ cases in the testing pool, $171$ cases are predicted correctly, i.e., $95.0\%$. To check if the prediction is biased, the predictions are categorized by the type of image. Table \ref{tb-results-one-qp} shows the number of times the actual image is $0$ ($1$) and the percentage of time QP of Fig. \ref{fig-5-1} predicted the digit correctly/incorrectly. 
\begin{table}[htbp]
  \centering
  \begin{tabular}{|c|c|c|c|}
    \hline
    \multicolumn{2}{|c|}{Total = } &\multicolumn{2}{|c|}{QP Prediction} \\\cline{3-4}
     \multicolumn{2}{|c|}{\newline 92 + 88} & 
     \quad {\bf 1} & \quad {\bf 0} \\ \hline
     \multirow{2}{*}{Data}  
     & \quad{\bf 1} \quad
     & \quad 91\% \quad & \quad 8.7\% \quad
     \\ \cline{2-4} 
     & \quad{\bf 0} \quad
     &  \quad 1.1\% \quad & \quad  98.9\% \quad
     \\ \hline
  \end{tabular}
  \caption{Success and failure prediction rates of the QP model (Fig. \ref{fig-5-1}) applied to the hand-written test data set for digits 0 and 1. The results show that the modeled QP is slightly biased toward predicting 0.}
  \label{tb-results-one-qp}
\end{table}

We performed the training and testing details five times and recorded a prediction accuracy on the test data set from $80.0\%$ to $95.0\%$. As one can see the two classes are almost balanced and the F1 score corresponding to the results shown in Table \ref{tb-results-one-qp} is $0.96$.


\subsection{QNN with two stabilizer codes}

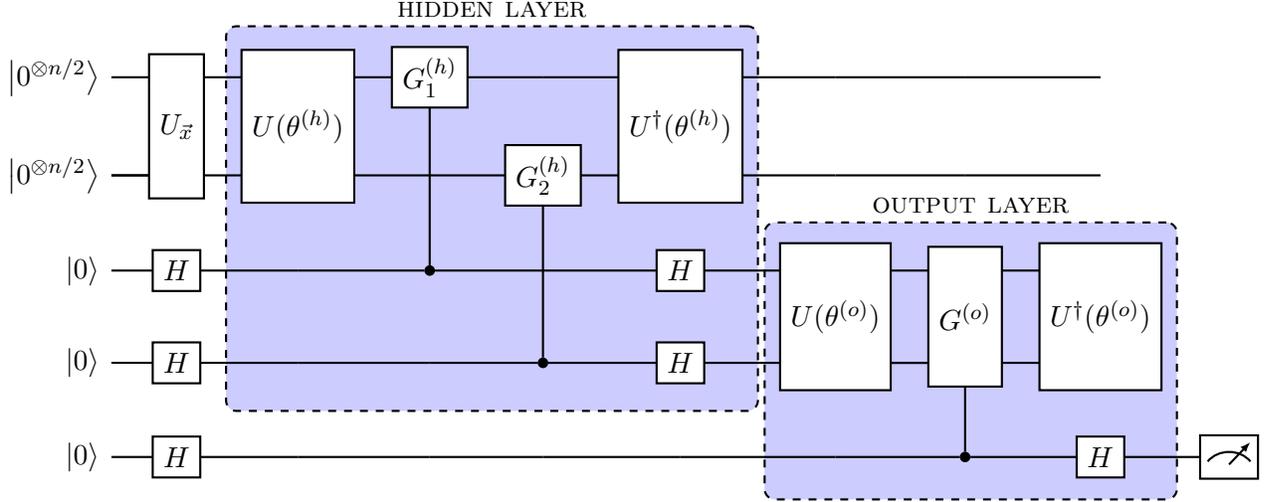
\begin{figure*}
\begin{quantikz}
\lstick{$\ket{0^{\otimes n/2}}$} 
&\gate[2]{U_{\vec{x}}}
& \gate[2]{U(\theta^{(h)})}
 \gategroup[4,steps=4,style={dashed,
                    rounded corners,fill=blue!20, inner xsep=2pt},
                    background]{{\sc hidden layer}}
& \gate{G^{(h)}_1}
& \qw
& \gate[2]{U^\dagger(\theta^{(h)})}
& \qw
& \qw
& \qw
\\
\lstick{$\ket{0^{\otimes n/2}}$} 
& \qw
& 
& \qw
& \gate{G^{(h)}_2}
& 
& \qw
& \qw
& \qw
\\
\lstick{$\ket{0}$} 
& \gate{H}
& \qw
& \ctrl{-2}
& \qw
& \gate{H}
& \gate[2]{U(\theta^{(o)})}
\gategroup[3,steps=3,style={dashed,
                   rounded corners,fill=blue!20, inner xsep=2pt},
                   background]{{\sc output layer}}
&  \gate[2]{G^{(o)}}
& \gate[2]{U^\dagger(\theta^{(o)})}
\\
\lstick{$\ket{0}$} 
& \gate{H}
& \qw
& \qw
& \ctrl{-2}
& \gate{H}
& 
& \qw
& 
\\
\lstick{$\ket{0}$} 
& \gate{H}
& \qw
& \qw
& \qw
& \qw
& \qw
& \ctrl{-1}
& \gate{H}
& \meter{}
\end{quantikz}
\caption{Example of a two-layer QNN. The input information $\ket{x}=U_{\vec{x}}\ket{\bf 0}$ is hypothesized to result from an unknown quantum encoding. The hidden and output layers are designed to interpret the input data by measuring a set of stabilizers. The hidden layer is equivalent of a stabilizer code defined (at $\theta^{(h)}={\bf 0}$) by two commuting generators $G_1^{(h)}$ and $G_2^{(h)}$. The outcome becomes the input of a secondary layer, a final stabilizer code with one generator $G^{(o)}$. To allow variation and exploration of other possible choices for the stabilizers, the parametric components $U(\theta^{(h)})$ and $U(\theta^{(o)})$ are introduced. 
}
\label{fig-5-2}
\end{figure*}
As another, more elaborate illustration of our proposal, consider the QNN in Fig. \ref{fig-5-2}, which comprises a \emph{hidden} layer with two QPs and an output layer with one QP. The first stabilizer code with two generators $G_1^{(h)}$ and $G_2^{(h)}$ (the hidden QPs) is used for the initial labeling of the input data. The output syndromes of the first quantum codes become the encoded data to a secondary stabilizer code (the output layer). To approximate the true encoding protocol of the incoming inputs $\ket{x}$, a parametric segment is added that probes for other potential generators as $G_k (\theta^{h}) = U^{\dagger}(\theta^{h})\, G_k^{(h)}\,U(\theta^{h})$, with $k=1,2$.
In the numerical studies, we adopt $G_1^{(h)}= \prod_{q=1}^{n/2} Z_q$ 
and  $G_2^{(h)}= \prod_{q=n/2+1}^{q=n} Z_q$. As in the previous case,  $n=6$. 
Notice that $G_1 (\theta^{h})$ and $G_2 (\theta^{h})$ commute for all possible parameters $\theta^{(h)}$, i.e., $\mathcal{S}_1 = \langle G_1 (\theta^{h}), G_2 (\theta^{h}) \rangle$, where $\theta^{(h)}$ is short hand for a set of parameters. 
The unitary operator $U(\theta^{(h)})  = \prod_{q=1}^{q=n} R_y( \theta^{(h)}_q)$ is chosen for this study. The output layer employs the stabilizer code with a single generator $G^{(o)}= Z_{n+1} Z_{n+2}$, and the variable parametric segment is $U(\theta^{(o)})  = R_y(\theta^{o}_{n+1}) R_y(\theta^{o}_{n+2})$. 

The input data, i.e., the training and testing subsets, are chosen to be the same as the single QP case in Fig. \ref{fig-5-1}. We trained the two-layer QNN of Fig. \ref{fig-5-2} five times and recorded the testing accuracy from $90.0\%$ to $97.8\%$. Generally, models with more parameters take more time to train.
%
We trained the model for $48$ seconds with the hope that circuits with more parameters will indeed require more time for training.
%
%

\emph{Results.-}
 Out of $180$ cases in the testing pool, $176$ cases are predicted correctly, i.e., $97.8\%$. To check if the prediction is biased the predictions are categorized by the type of image. Table \ref{tb-results-two-qp} shows the number of times the actual image is $0$ ($1$) and the percentage of time QNN predicted the digit correctly/incorrectly. The F1 score of this binary classification model is $0.98$.

\begin{table}[htbp]
  \centering
  \begin{tabular}{|c|c|c|c|}
    \hline
    \multicolumn{2}{|c|}{Total = } &\multicolumn{2}{|c|}{Two layer QNN Prediction} \\\cline{3-4}
     \multicolumn{2}{|c|}{92 + 88} & 
     \quad {\bf 1} & \quad {\bf 0} \\ \hline
     \multirow{2}{*}{Data}  
     & \quad{\bf 1\;} \quad
      & \quad \;\;\;\;\;95.7\%\;\;\;\;\;\;\;\quad & \quad 4.3\% \quad
     \\ \cline{2-4} 
     & \quad{\bf 0} \quad
      &  \quad 0\%\; \quad & \quad  100\% \quad
     \\ \hline
  \end{tabular}
  \caption{Table similar to \ref{tb-results-one-qp} but for the QNN model (Fig. \ref{fig-5-2}). The results show that the modeled QNN is slightly biased toward predicting 0. A scatter plot of this result is given in Fig. \ref{fig-5-3}.}
  \label{tb-results-two-qp}
\end{table}

\begin{figure*}
\includegraphics[width=\textwidth]{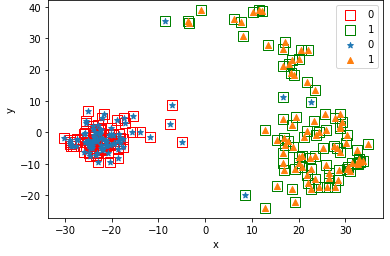}
\caption{PCA embedding of validation images. Each pixel value of an image can be considered as a dimension. The handwritten digit dataset has $64$ dimensions for each example, we reduced the dimensions of the validation images from $64$ to $2$ by using the PCA. Each data point is an image which is represented by the first $2$ principal components labeled as $x$ and $y$. Here, $\Box$ refer to the actual labels, while $\ast$ and $\triangle$ represent the predictions made during the validation process of the two-layer QNN model as shown in Fig. \ref{fig-5-2}.}
\label{fig-5-3}
\end{figure*}

Figure \ref{fig-5-3} shows the classification of the images by the corresponding predicted/actual label. We use Principal Component Analysis (PCA) to show the image data set in two dimensions, which correspond to the first two principal components.

\section{Quantum State Preparation - Exact Amplitude Encoding}
\label{sec-QSP}

The objective of the amplitude encoding (AE) is to represent classical data $\vec{x}$ as a normalized quantum state $\ket{x}$. The standard algorithm for providing an exact solution to this task is thoroughly explained in Ref. \cite{Shende2006} and implemented numerically in Ref. \cite{Qiskit}. 
The following review presents the steps of the exact algorithm from a slightly different angle. This different angle allows us to naturally introduce our approximate algorithm.
In the following sections we may refer to the amplitude encoding algorithm of
Ref. \cite{Shende2006} as the exact amplitude encoding. 

\emph{Problem Statement.--} 
An input  is assumed to be a vector 
$\vec{x} = [x_0, \cdots, x_{t-1}]^T$.
Without loss of generality, in the following we assume $t = 2^n$. 
In practice, if $t\ne 2^n$ for an integer $n$, we append $\epsilon \ne 0$ entries to $\vec{x}$ until the condition $t = 2^n$ is satisfied for the smallest possible integer $n$. For example, if 
$
\vec{x} = 
\left[
x_0, x_1, x_2 
\right]^T
$, 
with 
$t=3$,  
one can add a small  value 
$\epsilon \ne 0$ 
such that 
$
\vec{x} = 
\left[
x_0, x_1, x_2, \epsilon 
\right]^T
$
with  
$t=2^2$. 
While we will only consider real data, $x_k\in \mathbb{R}$, a generalization to 
complex data is straightforward. A real-world example of classical data vector $\vec{x}$ are the values of the pixels of an image. 

Given an input vector $\vec{x}$ with the properties described above, the goal is to initialize $n$ qubits in the quantum state
\begin{eqnarray}
\ket{x} = 
\frac{1}{\sqrt{\sum^{k=2^n-1}_{k=0} x_k^2}}
\,\,
\sum_j x_j \ket{j},
\label{eq-input-x}
\end{eqnarray}
where $\ket{j}$ stands for the usual bit string $\ket{j_n\cdots j_1}$, namely, $j_n\cdots j_1$ is the binary representation of integer $j$. 

\emph{Amplitude Encoding Core Step.--} 
Notice that $\ket{x}$ can be considered a matrix of dimensions $2^n \times 1$ ($2^n$ rows and 1 column), i.e., a column vector. This vector consists of two equal segments: upper and lower. Each segment is an unnormalized vector of 
with $2^{n-1}$ components. The upper segment contains the entries of $\ket{x}$ with $j_n=0$, and the lower segment contains the entries of $\ket{x}$ with $j_n=1$.

Suppose the existence of an $(n-1)$-qubit unitary operator 
$U^{(n-1)} = U^{(n-1)}(x_0, \cdots, x_{2^{n-1}-1})$ which is a function of the entries of the upper-half segment and prepares the quantum state 
%
\begin{eqnarray}
\ket{x^{u}} &=& 
\frac{1}{\sqrt{\sum^{2^{n-1} - 1}_{k=0} x_k^2}}
\,\,
\sum_j x_j \ket{j}
\nonumber \\
&=&  
U^{(n-1)} \ket{0^{\otimes (n-1)}}.
\label{eq-ket-up-x}
\end{eqnarray}
%
Similarly, suppose the existence of another $(n-1)$-qubit unitary operator 
$V^{(n-1)} = V^{(n-1)} (x_{2^{n-1}}, \cdots, x_{2^n-1})$ which is a function of the lower-half entries of $\vec{x}$ and prepares the quantum state 
%
\begin{eqnarray}
\ket{x^{d}} &=& 
\frac{1} {\sqrt{\sum^{2^n-1}_{k=2^{n-1}} x_k^2}}
\,\,
\sum_j x_j \ket{j}
\nonumber \\
&=&  
V^{(n-1)} \ket{0^{\otimes (n-1)}}.
\label{eq-ket-lower-x}
\end{eqnarray}
%
With the above assumptions and definitions, the state $\ket{x}$ can be expressed as 
\begin{eqnarray}
\ket{x} &=& 
\cos{(\lambda_n)}
\,
\ket{0_n} \ket{x^{u}} 
+
\sin{(\lambda_n)}
\,
\ket{1_n} \ket{x^{d}}
\nonumber \\
&=& 
\cos{(\lambda_n)}
\,
\ket{0_n} 
\left[
U^{(n-1)} \ket{0^{\otimes (n-1)}}
\right] \nonumber \\
&+&
\sin{(\lambda_n)}
\,
\ket{1_n}
\left[
V^{(n-1)} \ket{0^{\otimes (n-1)}}
\right]
,
\label{eq-x-decomp}
\end{eqnarray}
which represents the entanglement between qubit $n$ and the remaining $(n-1)$-qubit system. Notice that $\bra{x^{u}} \ket{x^{d}} \ne 0$ necessarily, i.e., Eq. (\ref{eq-x-decomp}) is not a Schmidt decomposition. 
Here, $\lambda_n$ is defined as
\begin{eqnarray}
\lambda_n
= 
\arccos{\left(\frac{N_{u}}{\sqrt{N_u^2 + N_d^2}}\right)}
\label{eq-x-2-lambda}
\end{eqnarray}
and $N_u$ and $N_d$ are the normalization factors of the upper-half and lower-half segments, respectively:
\begin{eqnarray}
 N_{u} &=& \sqrt{ 
 \sum_{k=0}^{2^{n-1}-1} \,\, x_k^2
 },
\end{eqnarray}
and
\begin{eqnarray}
 N_{d} &=& \sqrt{ 
 \sum_{k=2^{n-1}}^{2^n-1} \,\,x_k^2
 }.
\end{eqnarray}
The $\cos{(\lambda_n)}$ and $\sin{(\lambda_n)}$ factors in Eq. (\ref{eq-x-decomp}) ensure the proper normalization of the quantum sate $\ket{x}$.

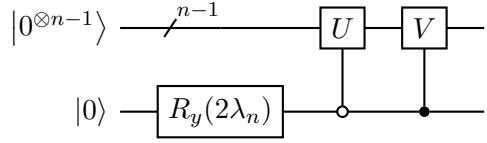
\begin{figure*}
\begin{quantikz}
\lstick{$\ket{0^{\otimes n-1}}$} 
& \qwbundle{n-1}
&\gate{U}
& \gate{V}
& \qw
\\
\lstick{$\ket{0}$} 
& \gate{R_y(2\lambda_n)}
& \octrl{-1}
& \ctrl{-1}
& \qw
\end{quantikz}
\caption{
Circuit representation of the AE algorithm:
$U^{(n-1)}$ prepares the normalized state corresponding to the upper-half of the input $\vec{x}$ [Eq. (\ref{eq-ket-up-x})], while
$V^{(n-1)}$ prepares the normalized state corresponding to the lower-half of the input $\vec{x}$ [Eq. (\ref{eq-ket-lower-x})].
The normalized state $\ket{x}$ is prepared applying the quantum circuit shown in the diagram; see Eq. (\ref{eq-x-decomp}). 
}
\label{fig-2-1}
\end{figure*}

Let us recall the definition of the $R_y$ single-qubit rotation gate in terms of the Pauli matrix $Y$,
\begin{eqnarray}
R_y(2\,\lambda)\ket{0} &=& 
e^{i Y \lambda}
\ket{0} 
\nonumber \\
&=&\cos{(\lambda)} \ket{0}
+
\sin{(\lambda)} \ket{1}.
\end{eqnarray}
Using $R_y$, the quantum state $\ket{x}$
in Eq. (\ref{eq-x-decomp}) can be prepared with the quantum circuit shown in Fig. \ref{fig-2-1}.

Similar to the decomposition of $\ket{x}$ into 
$\ket{x^u}$ and $\ket{x^d}$, 
each upper and lower segment can be further (and separately) decomposed.
This results in a recursive procedure that terminates when 
$\ket{x^u}$ and $\ket{x^d}$ correspond to single-qubit quantum states. 
The AE algorithm is based on the core step summarized in Fig. \ref{fig-2-1}. 
Details are provided in Appendix \ref{sec:appendix1}.

\emph{Circuit Depth.--} The depth of the quantum state preparation circuit described above is  $\mathcal{O}(2^n)$. Generally, the quantum state preparation circuit amounts to a decomposition of a $2^n \times 2^n$ unitary operator into one- and two-qubit gates \cite{Barenco1995PhysRevA523457,Reck1994PhysRevLett7358}. A simple explanation is that AE algorithm encodes segments of $2 \times 1$ dimension with multi-qubit gates, and there are exponential number of these segments in $\ket{x}$. The exponential depth necessitates the search for an approximate approach which we pursue in the next section.

\section{Approximating Amplitude Encoding}
\label{sec-QSP-approx}

As reviewed in Sec. \ref{sec-QSP}, the preparation of an arbitrary $n$-qubit state requires a $\mathcal{O}(2^n)$-depth circuit. The objective of this section is to introduce an approximate scheme that reduces the circuit depth to $\mathcal{O}(poly(n))$. 

\subsection{Recursive Approximate Scheme}
\label{sec:recursive_approx}

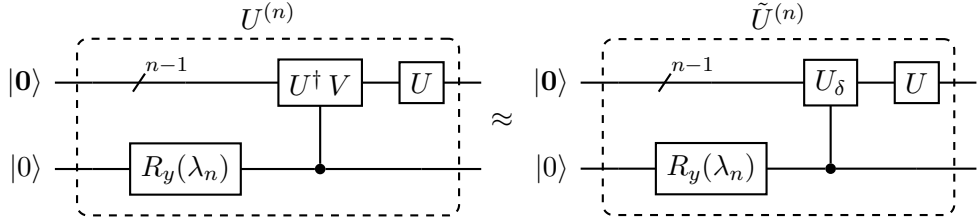
\begin{figure*}
\begin{quantikz}
\lstick{$\ket{{\bf 0}}$} 
&\qw
\gategroup[2,steps=4,style={dashed,
                   rounded corners, inner xsep=2pt},
                   background]{{\sc $U^{(n)}$}}
& \qwbundle{n-1}
&\gate{U^\dagger \,V}
& \gate{U}
& \qw
\\
\lstick{$\ket{0}$} 
&\qw
& \gate{R_y(\lambda_n)}
& \ctrl{-1}
& \qw
& \qw
\end{quantikz}
$\approx$
\begin{quantikz}
\lstick{$\ket{{\bf 0}}$} 
&\qw
\gategroup[2,steps=4,style={dashed,
                   rounded corners, inner xsep=2pt},
                   background]{{\sc $\tilde{U}^{(n)}$}}
& \qwbundle{n-1}
&\gate{U_{\delta} }
& \gate{U}
& \qw
\\
\lstick{$\ket{0}$} 
&\qw
& \gate{R_y(\lambda_n)}
& \ctrl{-1}
& \qw
& \qw
\end{quantikz}
\caption{An alternative, approximate implementation of amplitude encoding. A state $\ket{x^{(n)}}$ is encoded as $\ket{x^{(n)}} = U^{(n)} \ket{0}$. The state $U^\dagger \, V \ket{0}$ on the left-hand side circuit (which is exactly equivalent to the circuit in Fig. \ref{fig-2-1}) is approximated by $U_\delta \ket{0}$ on the right-hand side circuit, such that  $\ket{x^{(n)}} \approx \tilde{U}^{(n)} \ket{0}$. The approximation can be implemented recursively, where each $U^{(n-1)}$ and $V^{(n-1)}$ are approximations obtained from the previous iteration. The end result is a circuit of depth of $\mathcal{O}(poly(n))$.
}
\label{fig-3-1}
\end{figure*}

To see how the circuit depth can be reduced to a polynomial one, consider the basic idea behind the AE algorithm illustrated in Fig. \ref{fig-2-1} for an arbitrary number of qubits $q$. One can write the corresponding unitary evolution represented by this circuit as
\begin{equation}
    \ket{x^{(q)}} = U^{(q)}\,\ket{0^{\otimes (q)}} ,
\end{equation}
where
\begin{equation}
\label{eq-x-decomp-2}
  U^{(q)} =   \left[ \mathbb{I} \otimes U^{(q-1)} \right] \Bigl[
\ket{0_q}\bra{0_q} \otimes \mathbb{I}
+
\ket{1_q}\bra{1_q} \otimes
\left[ U^{(q-1)}\right]^{\dagger} \,\, V^{(q-1)}  
\Bigr] \left[ R_y(\lambda_q)\otimes \mathbb{I} \right] .
\end{equation}
This is  shown on the left-hand side of Fig. \ref{fig-3-1} for $q=n$. 
The quantum circuit on the left-hand side of Fig. \ref{fig-3-1} and the one in Fig. \ref{fig-2-1} are equivalent but the former, with a refactorized $U$ operator, is better suited for the approximation we discuss below.

While the composed operator $\left[U^{(q-1)}\right]^{\dagger} V^{(q-1)}$ cannot be easily approximated in general, the state $[U^{(q-1)}]^{\dagger} V^{(q-1)} \, \ket{0^{\otimes (q-1)}}$ can, up to a desired level of accuracy. The approximate scheme we proposed is based on the modified quantum-imaginary-time (MQITE) algorithm of Ref. \cite{Jouzdani2022}. Suppose that the state $[U^{(q-1)}]^{\dagger} V^{(q-1)} \, \ket{0^{\otimes (q-1)}}$ is approximated by $U_{\delta}^{(q-1)} \, \ket{0^{\otimes (q-1)}}$.
The details of this approximation are provided in the subsequent subsections. 
In this case, Eq. (\ref{eq-x-decomp-2}) can be rewritten as
\begin{eqnarray}
\ket{x^{(q)}} & \approx & 
U^{(q-1)} \Bigl[
R_{00}\ket{0} 
\, 
\ket{0^{\otimes (q-1)}}
+
R_{10}
\ket{1}
\,
U^{(q-1)}_{\delta}
\ket{0^{\otimes (q-1)}}
\Bigr] 
\nonumber \\
& \approx &
\tilde{U}^{(q)}
\,
\ket{0^{\otimes (q)}},
\label{eq-x-decomp-2-a}
\end{eqnarray}
where $R_{ij}=\bra{i_q} R_y(2\lambda_q)\ket{j_q}$ is the matrix element of the $Y$ rotation on the $q$-th qubit. An implementation of this approximation is shown on the right-hand side of Fig. \ref{fig-3-1}. 
Let $d[A^{(q)}]$ denote the depth of the part of the circuit associated to an operator $A$ acting on $q$ qubits. We can then relate the circuit depth of $\tilde{U}^{(q)}$ to the the circuit depths of the other operators in Fig. \ref{fig-3-1}, namely,
\begin{equation}
d[\tilde{U}^{(q)}] = d[R_y^{(1)}] + d[U_\delta^{(q-1)}] + d[U^{(q-1)}].
\end{equation}
Substituting $U^{(q-1)}$ by $\tilde{U}^{(q-1)}$ yields the recursive relation
\begin{equation}
d[\tilde{U}^{(q)}] = d[R_y^{(1)}] + d[U_\delta^{(q-1)}] + d[\tilde{U}^{(q-1)}].
\end{equation}
Setting $\tilde{U}^{(q_{\rm in}-1)} = U^{(q_{\rm in}-1)}$ for some initial number of qubits $q_{\rm in}$ allows one to arrive at
\begin{eqnarray}
d[\tilde{U}^{(n)}] &=& (n-q_{\rm in}+1)\, d[R_y^{(1)}] 
\nonumber \\
& & +\ \sum_{q=q_{\rm in}}^{n} d[U_\delta^{(q-1)}] + d[U^{(q_{\rm in}-1)}].
\label{eq-recursv-depth-a}
\end{eqnarray}
In practice, this recursive relation means that at every step $q$ of the approximate scheme, a block of data with $2^{q}$ entries is divided into upper and lower segments. Each segment is prepared with two unitary operators, where each of these operators is independently approximated as $\tilde{U}^{(q)}$ and $\tilde{V}^{(q)}$. Initially, one starts with a small integer $q_{\rm in}$ where the exact implementation of $U^{(q_{\rm in}-1)}$ and $V^{(q_{\rm in}-1)}$ is simple (e.g., $q_{in}=2$), and iterate the approximation steps until $q=n$. In addition, let us enforce $d[U_\delta^{(q-1)}] \sim \mathcal{O}(q^\alpha)$ at every step of the recursive iteration, where $\alpha$ is a positive power. Since both $d[R_y^{(1)}]$ and $d[U^{(q_{\rm in}-1)}]$ do not scale with $q$ (i.e., are $\mathcal{O}(1)$), we then have
\begin{equation}
d[\tilde{U}^{(n)}] \sim \mathcal{O}(poly(n)).
\label{eq-recursv-depth-b}
\end{equation}
Notice that $U^{(q_{\rm in}-1)}$ adds only a constant to the total circuit depth while the rotation operators $R_y^{(1)}$ combined generate a contribution of order $\mathcal{O}(n)$. Both contributions are subleading to the contribution coming from the $U_\delta$ operators introduced along the recursion.

\emph{Recursive Approximate-Scheme Algorithm.--} A formal description of the recursive approximate-scheme algorithm (RASA) is provided in Fig. \ref{fig:main-approx-algo}. The algorithm has parameters $\alpha$, $q_{\rm in}$, and the input data $\vec{x}$ as main inputs. It implicitly assumes that $\vec{x}$ complies with the conditions outlined in Sec. \ref{sec-QSP} and in Appendix \ref{sec:appendix1}, such as the length of $\vec{x}$ being $2^n$ for some integer $n$. Additionally, whenever $x_k=0, \forall x_k\in \vec{x}$, a small value $x_k\gets \epsilon$ is applied. The reason is that every segment of the data is to be encoded as a quantum state. If, for a segment, all values of $x_k$ are zero then the state has a normalization equal to zero, which causes a divide-by-zero error. 

In addition to the above inputs, $\chi$, the number of circuit executions, or circuit calls, must be specified. The number $\chi$ is based on the expected precision in the measurements. If $p$ significant figures are expected, then the number of circuit calls is $\mathcal{O}(10^{2p})$. 
Let $\eta$ be the number of components present in $\left[U\right]^{\dagger} V \ket{0^{\otimes(q)}}$ at a given step with $q$ qubits. The value of $\eta$ affects the precision of the measurements as well. When there are $\eta=q^{\alpha}$ components, the number of circuit calls to achieve $p$ significant figures is $\chi \ge q^\alpha \,\, 10^{2p}$.

Lines $2$--$4$ of the algorithm in Fig. \ref{fig:main-approx-algo} show $\alpha$, $p$, $\vec{x}$, and the initial number of qubits $q_{\rm in}$ as inputs ($n$ is specified by the bit-wise length of input data $\vec{x}$). 
On line $5$ of the algorithm, 
through the use of an external function {\bf PrepareInput},
the initial inputs are divided into blocks and represented as a collection (array) $S_{\ket{x}}$ of quantum states. 
%
On line $6$, {\bf InitialUnitaries} presents every state in $S_{\ket{x}}$  with an equivalent unitary quantum circuit. The corresponding circuit instructions are stored in the array $S_{U}^{(q_{\rm in})}$.
The recursion begins on line $7$ with $q$ taking values from 
$ q=q_{\rm in}$ to $q=n-1$.
For every value of $q$, 
a block in 
$S_{U}^{(q)}$ is merged to form a segment in a block of $S_{U}^{(q+1)}$.
The approximated circuits are named $U$ or $V$ depending on the segment they are referring to, i.e., the upper or lower segment. This is addressed on lines $11$--$14$: A subroutine {\bf Fuse} approximates and merges the two lower circuits with $q-1$ qubits into a larger one with $q$ qubits. 

It should be noted that in practice the inner loop of lines $10$--$16$ can be performed in parallel on separate quantum processors. 
A detailed description of the subroutines used in RASA is provided in Appendix \ref{sec:appendix3}.
An explicit example of this approximation scheme is given in Appendix \ref{sec:appendix2}.

\algblockx[Main]{StartP}{End}{\textbf{program Main}}{\textbf{end program}}

\begin{figure}    
\begin{algorithm}[H]
\caption*{Recursive Approximate-Scheme Algorithm}
\begin{algorithmic}[1]
\StartP
    \State with $\alpha, p \in \mathbb{Z}^+$ 
    \State with $\vec{x} = \left[x_0, \dots, x_{2^n-1}\right], \quad x_k \in \mathbb{R}$.
    \State with $q_{\rm in} \in \mathbb{Z}^+$ ; $1<q_{\rm in}<n$.     
    \State $S_{\ket{x}} \gets$ \textbf{PrepareInput}($\vec{x}$, $q_{\rm in}$)
    \State $S^{(q_{\rm in})}_{U} \gets$ \textbf{InitialUnitaries}($S_{\ket{x}}, q_{\rm in}$)
    \For{$q$ \textbf{in} $\left[q_{\rm in}, \cdots,  n-1\right]$ }:         
        \State $\chi=\mathcal{O}(q^\alpha  \,  \, 10^{2p})$ 
        \State with $S^{(q+1)}_{U} = \left[ \, \right]$ 
        \For{$b^\prime$ \textbf{in} $\left[ 0, \cdots, \frac{\vert \, S^{(q)}_{U} \, \vert}{2} -1\right]$}: 
            \State $\left(U^{(q-1)}, V^{(q-1)},N^U, N^V\right) \gets S^{(q)}_{U}[2b^\prime]$           
            \State  $(U_{b^\prime}^{(q)}, N_{b^\prime}^U) \gets  
                \textbf{Fuse}(U^{(q-1)}, V^{(q-1)},N^U, N^V, \chi, \alpha, p) $
            \State $\left(U^{(q-1)}, V^{(q-1)},N^U, N^V \right) \gets S^{(q)}_{U}[2b^\prime+1]$           
            \State $(V_{b^\prime}^{(q)}, N_{b^\prime}^V) \gets  
                    \textbf{Fuse}(U^{(q-1)}, V^{(q-1)},N^U, N^V, \chi, \alpha, p) $                    
            \State Append $\left[\left(U_{b^\prime}^{(q)}, V_{b^\prime}^{(q)},N_{b^\prime}^U, N_{b^\prime}^U \right) \right]$ to $S^{(q+1)}_{U}$
        \EndFor        
    \EndFor
\State $\chi=\mathcal{O}(n^\alpha  \,  \, 10^{2p})$ 
\State   $\left( U_{0}^{(n-1)}, V_{0}^{(n-1)},N_{0}^U, N_{0}^V\right) \gets S^{(n)}_{U}[0]$  
\State   $(U^{(n)}, N^U) \gets \textbf{Fuse}(U_{0}^{(n-1)}, V_{0}^{(n-1)},N_{0}^U, N_{0}^U, \chi, \alpha, p) $ 
\State   $x_k  \approx \, N^U \times \bra{k} U^{(n)} \ket{0^{\otimes n}}, 
\quad \forall \,x_k \, \in \vec{x}$
\End
\end{algorithmic}
\end{algorithm}    
\caption{Recursive Approximate-Scheme Algorithm. }
\label{fig:main-approx-algo}
\end{figure}


\subsection{Approximating $U^\dagger \, V \ket{ \bf 0}$}
\label{sec:approximate_U}

One of the tasks in RASA  is to merge two unitary operators $U^{(q-1)}$ and $V^{(q-1)}$ into one and obtain an approximation for $U^{(q)}$ (or $V^{(q)}$).
%
%
The very first step of this task is to approximate  $\left[U^{(q-1)}\right]^\dagger \, V^{(q-1)} \ket{\bf 0}$ as $U^{(q-1)}_\delta \ket{\bf 0}$, where $\ket{\bf 0}$ denotes a state with all qubits set to zero. It requires a quantum circuit $U^\dagger \, V \ket{\bf 0}$ to be executed $\chi$ times. At each execution the qubits are measured in the $Z$ basis. Each measurement corresponds to the observation of a bit string $\ket{j}$. 
After $\chi$ executions, by gathering the number of occurrences of a given $\ket{j}$, $n_j$, a list  $\left[ (\ket{j}, n_j) \right] $ is obtained. 
After sorting,  the first $q^\alpha$ largest components  are kept and the rest discarded. 
At this point, $S=\{(\ket{j}, \vert c_j \vert)\}$ is identified where $\vert c_j \vert^2 = n_j/\chi$, rounded to $p$ significant figures. 

The collection $S$ implies that  $\ket{\psi_\delta} \equiv \sum_{k=1}^{k=\eta} c_{j_k}\ket{j_k}$ is an approximation to $\left[U^{(q-1)}\right]^\dagger \, V^{(q-1)} \ket{\bf 0}$. 
Notice that the observed $\eta = \vert S\vert $ may be less than the cutoff $q^\alpha$. A second round of normalization is generally required to ensure $\sum_j \vert c_j\vert^2=1$, and therefore the  $\vert c_j\vert $ in $S$ should be divided by the normalization if necessary.
The phase in the coefficient $c_j$ is yet to be determined. The complete program to obtain the real and imaginary parts of $c_j$ is given in Ref. \cite{Jouzdani2022} and requires further executions of similar quantum circuits. The overall number of circuit executions is $\mathcal{O}(\eta\chi)$.

At this stage, $\eta\sim{\cal O}(q^\alpha)$ components of $\ket{\psi_\delta}$ state, i.e., $\{c_j\}$, have been identified. 
The next step is to be prepare this state. This is done by identifying the unitary operator $U_{\delta}$ such that $\ket{\psi_\delta} = U_{\delta} \ket{\bf 0}$. The latter operator can be constructed using the exact amplitude encoding algorithm (AE algorithm) reviewed in Sec. \ref{sec-QSP}.

\emph{Accuracy of the approximation.--} The accuracy of RASA strongly depends on the statistical distribution of the $\{\vert c_j\vert\}$ amplitudes in $U^\dagger \, V\ket{\bf 0}$. A sharp distribution around a single $\ket{j}$ means that only a few components are relevant and a limited number of executions is sufficient to accurately approximate these amplitudes. 
A broad or non-trivial distribution instead means that a limited number of circuit calls will not result in the complete list of $\vert c_j\vert$ with the expected precision; i.e., the ratio $n_j/\chi$ is not a good approximation to $\vert c_j\vert^2$. 

The true distribution of $\vert c_j\vert$ is in general unknown. 
However, in most practical applications one expects the distribution to be sharply centered around just a few bit strings. The justification is the following. Suppose the data under consideration is an image and an entry in $\vec{x}$ corresponds to the intensity of a pixel of the image. 
We can interpret $U \ket{\bf 0}$ and $V \ket{\bf 0}$ as the encodings of two neighboring parts of the \emph{same} image. It is natural to assume that these two parts are correlated to some degree. $U^\dagger \, V \ket{ \bf 0}$ can be understood as looking at the contrast of the two neighboring parts of the same image. The correlation means that this quantum state, which is the back propagation of $V \ket{ \bf 0}$ by $U$, should intuitively be close to a single bit string $j$, or perhaps a limited number of them. 
An extreme scenario is an image with translational invariance. Then  $U \ket{\bf 0}$ and $V \ket{\bf 0}$ are replicas, perhaps up to a phase. Thus  $U^\dagger \, V \ket{ \bf 0} $ is equal to $\ket{ \bf 0}$ up to a phase. 
This argument supports the assumption that often the set $\{\vert c_j \vert \}$ is sharply distributed around a limited number of bit strings. 
Since $ \ket{\psi_\delta} \approx U^\dagger \, V \ket{ \bf 0}$ has only a few components, they can be identified with good accuracy by the limited number of circuit execution and measurements. In this scenario,
$U_\delta\ket{\bf 0}$ is constrained to be polynomial in the circuit depth. 

\emph{Correlation length.--} In most real-world scenarios, certain correlations between parts of data exist, although the correlations may be weak or strong at different scales (i.e., the size of the blocks). By adjusting the starting number of qubits, $q_{\rm in}$, one can adapt to the relevant correlation scales in the input data, and thus improve accuracy. 
%

\subsection{Relation to the MQITE Algorithm}
\label{sec-mqite}

Reference \cite{Jouzdani2022} introduced a similar approximate scheme where the \emph{contrast} between two states was assumed to be limited to a certain number of components. The objective of the MQITE algorithm of Ref. \cite{Jouzdani2022} was to estimate the ground state of a given Hamiltonian.  
Specifically to that case, at every step one had $V=Q_k\, U$, where $Q_k$ is a local unitary operator in the Hamiltonian $H=\sum_{k} w_k\, Q_k$ with corresponding weight $w_k$.
In MQITE, the assumption that $U^{\dagger}\,V\ket{\bf 0}$ has only a limited number of components emerged from the argument that  
$U\ket{\bf 0}$ is close to the ground state of 
$H$ and thus $U^{\dagger}\,V\ket{\bf 0}$ captures perturbations induced by $H$ on $U\ket{\bf 0}$.

An important feature of the MQITE algorithm is that the components of $\ket{\psi_{\delta}} = U_\delta \ket{\bf 0}\approx U^{\dagger}\,V\ket{\bf 0}$
can be measured using solely a quantum computer and one ancillary qubit,
provided that the assumption $\eta \sim poly(n)$ holds. No auxiliary classical computer is required.



\subsection{Numerical Validation}

\begin{figure}[h]
 \includegraphics[width=1\textwidth]{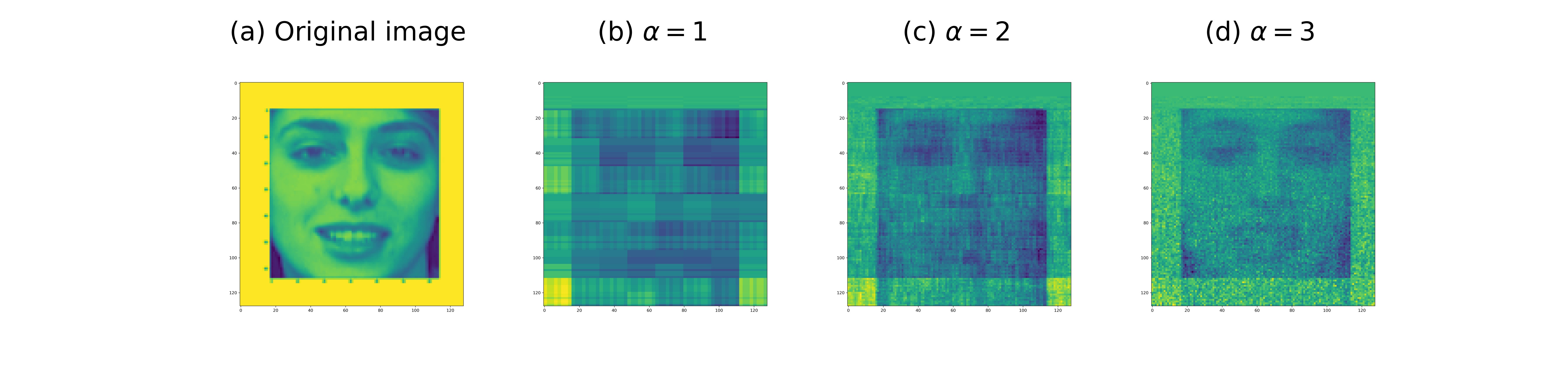}
\caption{
Comparison between original and encoded images when the parameter $\alpha$ is changed. 
(a) The original image \cite{scikit-learn}. 
(b) With $\alpha=1$
(c) For $\alpha=2$, the approximation begins at a larger scale
(d) For $\alpha=3$, the approximation begins at even larger blocks, and RASA finds meaningful correlations. 
 }
\label{fig-3-3}
\end{figure}

To illustrate and check the applicability of the recursive approximate scheme introduced above, we consider a $2^7 \times 2^7$ square image containing $16,384$ pixels, as shown in Fig. \ref{fig-3-3}(a).
Each  pixel has a value  $x\in\mathbb{R}^+$. The input is prepared as a $2^{14} \times 1$ vector $\vec{x}$. The objective is to load (encode) $\vec{x}$ into
$n=14$ qubits using RASA.

As an initial step, the input is divided into $2^{13}$ blocks of length $2$, i.e., $q_{\rm in}=2$.
The $2\times 1$ corresponding state vectors are normalized and the normalizations are recorded. Every $2\times 1$ normalized state is then represented as the state of a single qubit. The corresponding unitary operators are identified and their decompositions in terms of elementary gates ($S_{U}^{(2)}$ are recorded.
The recursive scheme is followed and the iterations continue until only one block remains.

The algorithm is implemented using the IBM quantum simulator \cite{Treinish2022Qiskit/qiskit:0.34.2}. 
In our numerical study, $\chi=4\times 10^4$ is fixed through all the iterations and $p=2$ is assumed. 
At every circuit run, the qubits are measured in the computational $Z$-basis, yielding a bit string $\ket{j}$. 
A record is kept of all $\chi$ resulting bit strings and the set $\{n_j\}$ is constructed, where $n_j$ is the number of times a bit string $j$ is obtained.
After $\chi$ executions, only the $q^\alpha$ most frequent bit strings are kept and the rest are  discarded. Then, $ \sqrt{n_j/\chi}$ is rounded to $p=2$ significant figures and is considered as an approximation to the true value of $\vert c_j \vert $. 
The amplitudes are further renormalized to ensure $\sum_j \vert c_j \vert^2 = 1$.
In our numerical study, we use the exact state vector of $U^\dagger V \ket{0}$ to obtain the sign of a $c_j$ component (when using quantum hardware, the sign is obtained via additional quantum circuit executions, see Ref. \cite{Jouzdani2022}). 

At every step and for every block, a set of $\{(\ket{j}, c_j$)\} bit strings and coefficients are determined, which means the vector $U^\dagger V \ket{0}$ is approximated as $\ket{\psi_\delta} = \sum_{k=1}^{k=\eta} c_{j_k} \ket{j_k}$.  
This state, which now has only $\mathcal{O}(poly(q))$ components,  can be prepared on $q-1$ qubits by using an algorithm for quantum state preparation. We use the exact quantum state preparation approach reviewed in Sec. \ref{sec-QSP}. 

Figures \ref{fig-3-3}(b)-(c) show the result of repeating the above procedure for different values of $\alpha$.  For example, in Fig. \ref{fig-3-3}(b), at each iteration step $q$, only the first $q$ most frequent bit strings $\ket{j}$ were selected (i.e., $\alpha=1$). 
This means that the state $\left[U^{(q-1)}\right]^\dagger \, V^{(q-1)} \ket{\bf 0}$
is approximated by a state with only $q$ components.
Similarly, in Figs. \ref{fig-3-3}(c) and \ref{fig-3-3}(d) the cutoffs are $q^2$ and $q^3$ respectively. 

The gradual improvement of the images can be understood as follows. 
Consider comparing the circuit depth of the two different approaches: the exact amplitude encoding and the recursive approximation scheme. 
By setting $d[R_y^{(1)}]=1$ and 
$d[U_\delta^{(q-1)}]=q^\alpha$, we have 
\begin{eqnarray}
d[\tilde{U}^{(n)}] &=& (n-q_{\rm in}+1)\,  
\nonumber \\
& & +\ \sum_{q=q_{\rm in}}^{n} q^{\alpha} + 2^{q_{\rm in}-1}
\label{eq-recursv-depth-recursive}
\end{eqnarray}
for the approximation and 
\begin{eqnarray}
d[U^{(n)}] &=& (n-q_{\rm in}+1)
\nonumber \\
& & +\ \sum_{q=q_{\rm in}}^{n} 2^{(q-1)}+ 2^{q_{\rm in}-1}
\label{eq-recursv-depth-exact}
\end{eqnarray}
for the exact approach.
Our goal is to find out up to what value of $q$ the two  approaches yield the same results. In other words, when the approximation is exact. This allows us to determine the scale at which the parts of the image were encoded exactly. If this scale is too small there may not be enough correlation and thus the approximation fails. Conversely, if the scale is large enough, such that the two segments in each block have considerable correlation, then the approximation should qualitatively converge to the original image. It can be verified from Eqs. (\ref{eq-recursv-depth-recursive}) and (\ref{eq-recursv-depth-exact}) that for $\alpha=1$, the threshold is $q=2$. At this scale, one segment contains two pixels each. 
The data are completely uncorrelated and thus the approximation fails. 
For $\alpha=2$, the threshold is approximately $q=3$. There is some improvement and the image of a face is emerging in Fig. \ref{fig-3-3}(c), although vaguely.
For $\alpha=3$ the threshold is approximately $q=11$ (see Fig. \ref{fig-3-3}(d)). At this scale, the segment contains enough amount of information. They are correlated to the adjacent segments. Once the correct scale is reached at $\alpha>2$, the encoded image has enough detailed information to be reasonably discernible.

\begin{figure}[h]
 \includegraphics[width=1\textwidth]{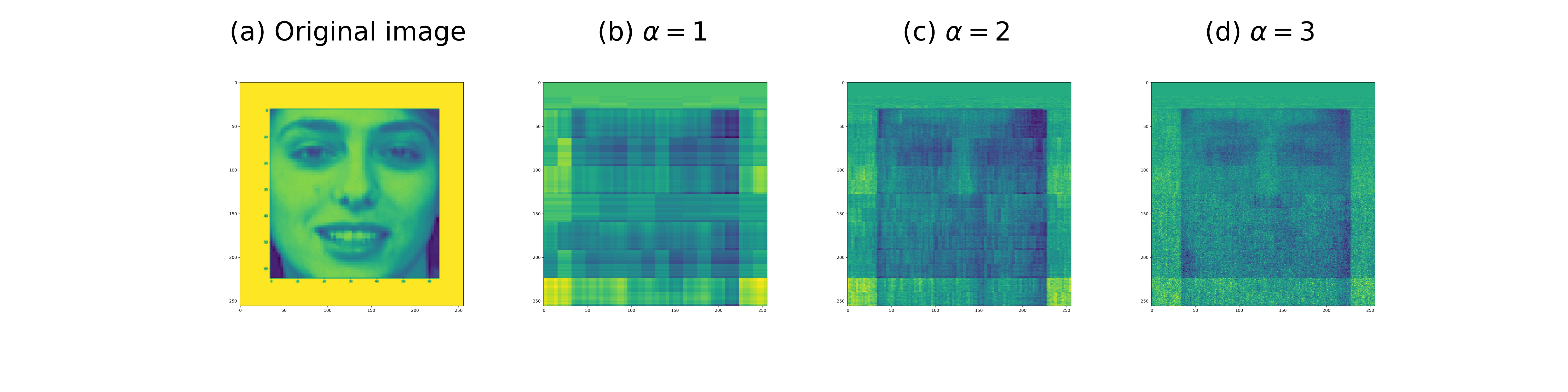 }
\caption{
Same as Fig. \ref{fig-3-3}, but with an original picture (a) that has finer resolution. Here the image has $2^8\times 2^8$ pixels. Nevertheless, resemblance to the original image begins to emerge in (c) and improves in (d) with $\alpha=3$. 
}
\label{fig-3-3-b}
\end{figure}

Figure \ref{fig-3-3-b} shows the results of the same numerical simulation as in Fig. \ref{fig-3-3} but performed with a grid of $2^8\times 2^8$ pixels. Although the image is finer, we observe qualitatively similar results. For $\alpha=3$, the segments with $q=11$ are coarse enough to allow the approximation to capture relationship between them. An image resembling closely the original one emerges.

\section{Summary and conclusion}
\label{sec-summary}


In the first part of this paper, a data classification protocol is introduced. This protocol borrows the concept of \emph{parity check} and \emph{stabilizer code} from the literature of quantum error-correcting codes to compose a quantum circuit module which represents a perceptron found in ANNs. Parity check allows one to classify input data. We employ this classification ability to mimic the behaviour of a perceptron. Furthermore, additional features could be extracted from the same data by using a collection of commuting stabilizers, each measuring the occurrence of a separate fictitious error. In this picture, a stabilizer code corresponds to a layer in an ANN. By passing the output of one code to another, we concatenate layers and create what resembles a deep ANN. We illustrate our protocol by classifying a dataset of hand-written digits.

It should also be pointed out that any stabilizer code is characterized by a code distance. The code distance is a metric  indicating the strength of a code in detecting and correcting different types of errors. Translated to QNNs as  in this paper, a code distance can potentially be used as a metric to quantify the data classification strength of QNNs, a subject future investigation.

In the second part of this paper, an approximate approach to load input data into quantum memory is introduced. 
The recursive approximate-scheme algorithm (RASA) uses contrast between parts of the input data to reduce the number of required gates for encoding. The underlying logic can be understood in the following comparison. Suppose the goal is to save a human face image that has $8 \times 8$ pixels. The standard approach is to send $64$ separate instructions to the machine for every pixel. However, knowing that a face is relatively symmetric, an alternative approach is to save only half of the the image ($32$ instructions), plus instructions for reconstructing the other half of the image from the first half. If the number of latter instructions is less than $32$, then we save in the number of instructions needed to encode the image. 

In this analogy, the number of instructions is equivalent to the circuit depth. We show that input data can be split in segments. By considering the correlation and contrast amongst these segments, we can reduce the circuit depth required to encode the input data in a quantum memory. In particular, the amount of detail shared between two segments can be set on-demand and therefore the depth of the circuit can be kept tractable. We further hypothesize that for most real-world inputs there are enough correlations to be used that allows one to constrain circuit depth to polynomial in the number of qubits. The feasibility of the circuit depth control and the resolution reinforcement is illustrated by numerical simulations.

Quantum algorithms that search for the ground or excited states of a give Hamiltonian \cite{tkachenko2022quantum, Gomes2021, stetcu2022projection, Huang2023} often require working with a parametric circuit, known as an ansatz. Some prior knowledge of the underlying symmetries in the ground state of the problem in hand, combined with the recursive approach introduced in this paper, can potentially allow one to express a ground state as a parameterized circuit with only polynomial depth. Further study in this direction is needed. 

All simulations in this paper are performed under assumption of a perfect quantum device, i.e., the effects of errors, noise, and decoherence were not considered. The impact of noisy gates and imperfect measurement on the main results of the paper also requires further study.

\section{Acknowledgments}
\label{sec-acknowledgments}
This material is based upon work supported by the U.S. Department of Energy, Office of Science, Office of Advanced Scientific Computing Research under Award Number DE-SC0023398.

This report was prepared as an account of work sponsored by an agency of the United States Government.  Neither the United States Government nor any agency thereof, nor any of their employees, makes any warranty, express or implied, or assumes any legal liability or responsibility for the accuracy, completeness, or usefulness of any information, apparatus, product, or process disclosed, or represents that its use would not infringe privately owned rights.  Reference herein to any specific commercial product, process, or service by trade name, trademark, manufacturer, or otherwise does not necessarily constitute or imply its endorsement, recommendation, or favoring by the United States Government or any agency thereof.  The views and opinions of authors expressed herein do not necessarily state or reflect those of the United States Government or any agency thereof.

We would like to specially thank GA team: Mark Kostuk, Christian Zuniga, and Matthew Cha, for the their inputs and feedbacks that helped us improve the  manuscript.

\clearpage
\appendix
\section{Detailed Review of Exact Amplitude Encoding Algorithm}
\label{sec:appendix1}

\subsection{Amplitude encoding by iteration from $n$ to $1$}
\label{sec-sec-general-case}

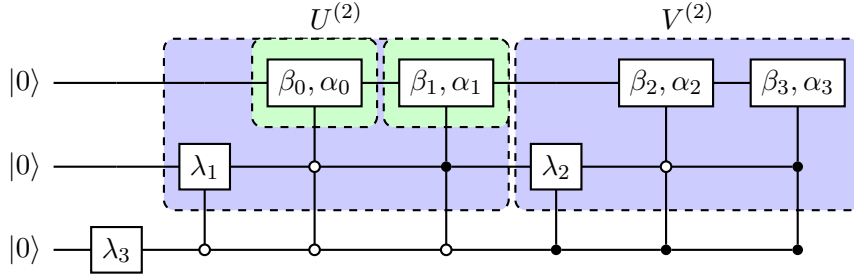
\begin{figure*}
\begin{quantikz}
\lstick{$\ket{0}$} 
& \qw
& \qw
\gategroup[2,steps=3,style={dashed,
                   rounded corners,fill=blue!20, inner xsep=2pt},
                   background]{$U^{(2)}$}
& \gate{\beta_0, \alpha_0}
\gategroup[1,steps=1,style={dashed,
                   rounded corners,fill=green!20, inner xsep=2pt},
                   background]{}
& \gate{\beta_1, \alpha_1}
\gategroup[1,steps=1,style={dashed,
                   rounded corners,fill=green!20, inner xsep=2pt},
                   background]{}
& \qw\gategroup[2,steps=3,style={dashed,
                   rounded corners,fill=blue!20, inner xsep=2pt},
                   background]{$V^{(2)}$}
& \gate{\beta_2, \alpha_2}
& \gate{\beta_3, \alpha_3}
\\
\lstick{$\ket{0}$} 
& \qw
& \gate{\lambda_1}
&  \octrl{-1} 
&  \ctrl{-1} 
&  \gate{\lambda_2}
&  \octrl{-1}
&  \ctrl{-1} 
\\
\lstick{$\ket{0}$} 
& \gate{\lambda_3}
& \octrl{-1}
& \octrl{-1}
& \octrl{-1}
& \ctrl{-1}
& \ctrl{-1}
& \ctrl{-1}
\end{quantikz}
\caption{
A quantum circuit to prepare an arbitrary (normalized) input data 
$\vec{x}=\left[x_0, x_1, \cdots, x_7\right]^T$ as a quantum state vector $\ket{x}$. 
The gate parameters are determined by the input data. 
The normalized state vector $\ket{x}$ 
can be seen as $2^{n-1}$ blocks, each block is $2\times 1$ in size. 
Each block is populated by a MQC gate that is  
controlled by qubits $2$ to $n$.
Example of targeted $2\times 1$ blocks are shown in green.  
Notice that each block is completely specified from an initial $\ket{0}=[1,0]^T$ state by a rotation $R(2\beta)$ with $\beta$ rotation angle, followed by  $Z^\alpha$ with $\alpha=0$ or $\alpha=1$. This is a special case of an SU(2) gate that would be applicable in case of complex vectors.
The rest of the MQC gates (with $\lambda$ parameters) adjusts the relative phase of the blocks. 
}
\label{fig-2-2}
\end{figure*}

An example of the recursive approach used in the exact amplitude encoding is shown in Fig. \ref{fig-2-2} for $n=3$.
The quantum circuit prepares any arbitrary input real quantum state $\ket{x}$ defined on $n=3$ qubits. 
This recursive process is the underlying concept introduced in Ref. \cite{Shende2006}. The circuit contains
an exponential number (in $n$) of multi-qubit control gates (MQC gates). 
Each MQC gate targets a $2\times 1$ block of the input $\vec{x}$. 
An MQC gate manipulates  the first qubit of the circuit and prepares it as a real $2\times 1$ vector. This is shown as a green box in Fig. \ref{fig-2-2}.
This manipulation is composed of two actions. 
Suppose the $2\times 1$ vector is $\left[x, x^\prime\right]^T$.
The first action is a rotation $R_y(2\beta)$ which turns $\ket{0}$ 
into 
$\cos{\beta} \ket{0} +\sin{\beta}\ket{1}$, such that 
$\cos{\beta}$ is proportional to $x/\sqrt{x^2 + {x^{\prime}}^2 }$ and $\sin{\beta}$ is proportional to $x^\prime/\sqrt{x^2 + {x^{\prime}}^2 }$.
The second action is to adjust the relative phase between 
$\cos{\beta} \ket{0}$ and $\sin{\beta} \ket{1}$ to match with the relative phase in the real-value input $\left[x, x^\prime\right]^T$. 
The phase is either $+1$ where no action is needed, or $-1$. The latter can be achieved by applying a $Z$ gate. 
The relative phase cannot be captured by a  rotation $R_y(2\beta)$ alone as it physically indicates if the two-dimensional coordinate system is right-handed or left-handed.
Let us compactly denote the second action as $R_\alpha = Z^\alpha$ where $R_0=I$ and $R_1=Z$.
Therefore, with proper variables, $R_\alpha R(2\beta)\ket{0}$ prepares a normalized version of a given real $2\times 1$ input, i.e., $(x\ket{0}+x^\prime \ket{1})/\sqrt{x^2 + {x^{\prime}}^2 }$.

Using Eq. (\ref{eq-x-2-lambda}), the components of $\lambda$ are computed from the relations
\begin{eqnarray}
 \sin{(\lambda_3)} = \frac{\sqrt{x_0^2 + \cdots + x_3^2 }}{\sqrt{x_0^2 + \cdots + x_7^2 }}
 \nonumber \\ 
 \sin{(\lambda_{1})} = \frac{\sqrt{x_0^2 + x_1^2 }}{\sqrt{x_0^2 + \cdots + x_3^2 }}
 \nonumber \\ 
 \sin{(\lambda_{2})} = \frac{\sqrt{x_4^2 + x_5^2 }}{\sqrt{x_4^2 + \cdots + x_7^2 }},
 \label{eq-apend-lambda}
\end{eqnarray}
and the $\beta$ parameters follow from
\begin{eqnarray}
 \sin{(\beta_{0})} = \frac{x_0}{\sqrt{x_0^2 + x_1^2 }}
 \nonumber \\ 
 \sin{(\beta_{1})} = \frac{x_2}{\sqrt{x_2^2 + x_3^2 }},
 \nonumber \\ 
 \sin{(\beta_{2})} = \frac{x_4}{\sqrt{x_4^2 + x_5^2 }},
 \nonumber \\ 
 \sin{(\beta_{3})} = \frac{x_6}{\sqrt{x_6^2 + x_7^2 }}.
 \label{eq-apend-beta}
\end{eqnarray}

Each box with parameter $\beta$ and $\alpha$ is an operation on a qubit that prepares a \emph{normalized} state. For example in Fig. \ref{fig-2-2}, inside $U^{(2)}$, the green boxes are associated with two normalized states $\ket{\alpha_0, \beta_0}$ and $\ket{\alpha_1, \beta_1}$. The role of MQC gates with a $\lambda$ parameter is to  adjust the normalization factor. For example in $U^{(2)}$, the  overall state is $\cos{\lambda_1} \ket{0}\ket{\alpha_0, \beta_1} + \sin{\lambda_1} \ket{1}\ket{\alpha_0, \beta_1}$. With the value of $\lambda_1$ defined in Eq. (\ref{eq-apend-lambda}), this state matches with the \emph{normalized} upper half of the input $\vec{x}$. Likewise, $\lambda_3$ is used to 
adjust the coefficients of $\ket{0}\left[U^{(2)}\ket{00}\right]$ and $\ket{1}\left[ V^{(2)}\ket{00}\right]$ to 
ensure the final output of the circuit is the normalized version of $\vec{x}$, i.e., $\ket{x}$.

\subsection{Preparing  the input}

For the problem defined in Sec. \ref{sec-QSP},
consider a situation where the input is $\vec{x} = \left[0, 0\right]$. This clearly results in a divide-by-zero error, for example in Eqs. (\ref{eq-apend-beta}). 
Now, consider an input $\vec{x}$ that has one or more $2\times 1$ blocks of $\left[0, 0\right]$. Since the normalization of these blocks is zero, it is impossible to encode the block with a single-qubit unitary operator acting on a qubit. 
To fix this situation, it is necessary to adjust the inputs. Specifically, adjusting means replacing every data entry $x_k$ with $f(x_k)$ where $f$ is a one-to-one function with known inverse $f^{-1}$. For example, when all entries $x_k$ are nonzero, $f: x_k \to x_k$ is sufficient. Another possible choice can be a sigmoid function $f(x) = [1+\exp(-\alpha x)]^{-1}$.
Adjusting or preparing input data (which amounts to a filtering) can have additional benefits. For example, the filtered data can have more contrast between different segments. This allows one to better capture the correlation between two segments which may not be significant in the original data.

\section{Subroutines used in RASA}
\label{sec:appendix2}

Figures \ref{fig:fuse-algo}, \ref{fig:approx-algo}, \ref{fig:PrepareInput}, and \ref{fig:InitialUnitaries} show the subroutines {\bf Fuse}, {\bf Approx}, {\bf PrepareInput}, and {\bf InitialUnitaries}, respectively, which are used in the recursive approximate algorithm.

\algblock[Name]{Function}{EndFunction}
\begin{figure}[h]
    \centering
\begin{algorithm}[H]
\begin{algorithmic}[1]
\Function{ \textbf{Fuse}}{ $U^{(q-1)},V^{(q-1)}, N_u, N_v, \chi, \alpha, p$ }
    \State $U^{(q-1)}_{\delta} \gets \textbf{Approx}( U^{(q-1)}, V^{(q-1)}, \chi, \alpha, p )$
    \State  $\lambda         \gets$    $(N_u, N_v)$ in Eq. (\ref{eq-x-2-lambda}) 
    \State  $\tilde{U}^{(q)} \gets$    $(\lambda, U^{(q-1)}_{\delta}, U^{(q-1)})$ in Eq. (\ref{eq-x-decomp-2-a})
    \State \textbf{Return} ($\tilde{U}^{(q)}$, $N$)
\EndFunction
\end{algorithmic}
\end{algorithm}
    \caption{Fusing the two segments of a block through approximation. $N=\sqrt{N_u^2+N_v^2}$ is the normalization factor associated to $\tilde{U}^{(q)}$.}
    \label{fig:fuse-algo}
\end{figure}

\begin{figure}[h]
    \centering
\begin{algorithm}[H]
\begin{algorithmic}[1]
\Function{\textbf{Approx}}{ $U^{(q-1)}, V^{(q-1)}, \chi, \alpha, p $ }
    \For{$r=1$ \textbf{to} $\chi$} (on a quantum computer):     
        \State execute $\left[U^{(q-1)}\right]^\dagger \, \, V^{(q-1)} \,\, \ket{0^{\otimes(q-1)}}$  \Comment{See Ref. \cite{Jouzdani2022}}
        \State record observed $ \ket{j}$ \Comment{$\Leftrightarrow$ $n_j=n_j+1$}     
    \EndFor
    \State $S=\{ \left( \ket{j}, \vert{c_j}\vert  \right) \}$ 
    \Comment{$\vert{c_j}\vert=\frac{n_j}{\chi}$, sort, pick $q^\alpha$ largest entries}     
    \State   round $\vert{c_j}\vert \quad \forall \, \vert c_j \vert \in S$
    \Comment{to $p$ significant figure.}
    \State $\eta = \vert S \vert $ 
    \Comment{$\eta \le q^\alpha$ }
    \State $\{ c_j = c_j^{(r)} + ic_j^{(i)} \} \gets $ $S$ and Ref. \cite{Jouzdani2022}. 
    \State $\ket{\psi_\delta} = \sum_{k=1}^{k=\eta} c_{j_k} \ket{j_k} \gets \left[U^{(q-1)}\right]^\dagger \, \, V^{(q-1)} \,\, \ket{0^{\otimes(q-1)}}$.        
    \State \textbf{Construct} $U_\delta$ such that $\ket{\psi_\delta} = U_\delta\ket{0^{\otimes(q-1)}} $ 
    \Comment{f.e.x. using AE Algorithm }
    \State \textbf{Return} $U_\delta$ \Comment{ Notice $d[U_\delta]=\mathcal{O}(\eta) \le \mathcal{O}(q^\alpha$)}
\EndFunction
\end{algorithmic}
\end{algorithm}
    \caption{Approximating $U^\dagger V \ket{\bf 0}$ by execution on a quantum computer and finding the most dominant components. This subroutine 
    approximates $U^\dagger V \ket{\bf 0}$ as  $\ket{\psi_\delta}$ on line $10$.  The state can be prepared by $U_\delta\ket{\bf 0}$. The function returns the circuit instruction $U_\delta$. 
    }
    \label{fig:approx-algo}
\end{figure}


\algblock[Name]{Function}{EndFunction}
\begin{figure}[h]
    \centering
    \begin{algorithm}[H]
        \begin{algorithmic}[1]
            \Function{\textbf{PrepareInput}}{$\vec{x}, q$}
                \State $S_{\ket{x}}=\left[\, \right]$
                \State $n_b=\frac{\vert\vec{x}\vert}{2^{q}}$ and $t_i=t_f=0$
                \For{$b$ \textbf{in} $\left[0, \cdots, n_b-1 \right]$}:
                    \State $t_i=t_f$
                    \State  $t_f=t_i +  2^{q-1} $
                    \State $ x^u = \left[ x_{t_i} \cdots, x_{t_f-1} \right]$                    
                    \State $N_u \gets $ norm of $x^u$
                    \State $t_i=t_f$
                    \State $t_f=t_i+2^{q-1}$
                    \State $ x^d = \left[ x_{t_i}, \cdots, x_{t_f-1} \right]$ 
                    \State $N_d \gets $ norm of $x^d$
                    \State Append $\left[( \frac{x^u}{N_u},  \frac{x^d}{N_d}, N_u, N_d) \right]$ to $S_{\ket{x}}$
                \EndFor
                \State \textbf{Return} $S_{\ket{x}}$
            \EndFunction
        \end{algorithmic}
    \end{algorithm}
    \caption{PrepareInput subroutine.}
    \label{fig:PrepareInput}
\end{figure}

\begin{figure}[h]
    \centering
    \begin{algorithm}[H]
        \begin{algorithmic}[1]
            \Function{\textbf{InitialUnitaries}}{$S_{\ket{x}}, q$}        
                \State $S^{(q)}_{U}=[\,]$
                \For{$b$ \textbf{in} $\left[0, \cdots, \vert S_{\ket{x}} \vert - 1 \right]$}:
                    \State $(\ket{x^u} , \ket{x^d}, N_u, N_d) \gets S_{\ket{x}}[b]$
                    \State Find  $U^{(q-1)}$ such that $\ket{x^u} = U^{(q-1)} \, \ket{0^{\otimes (q-1)}}$  \Comment{ f.e.x. using AE Algorithm }
                    \State Find  $V^{(q-1)}$ such that $\ket{x^d} = V^{(q-1)}\ket{0^{\otimes (q-1)}}$                      
                    \State Append $\left[( U^{(q-1)},  V^{(q-1)}, N_u, N_d)\right]$ to $S^{(q)}_{U}$
                \EndFor  
                \State \textbf{Return} $S^{(q)}_{U}$
            \EndFunction
        \end{algorithmic}
    \end{algorithm}
    \caption{InitialUnitaries subroutine.}
    \label{fig:InitialUnitaries}
\end{figure}

\section{Recursive Approximation Example}
\label{sec:appendix3}

\begin{figure*}
\includegraphics[scale=0.5]{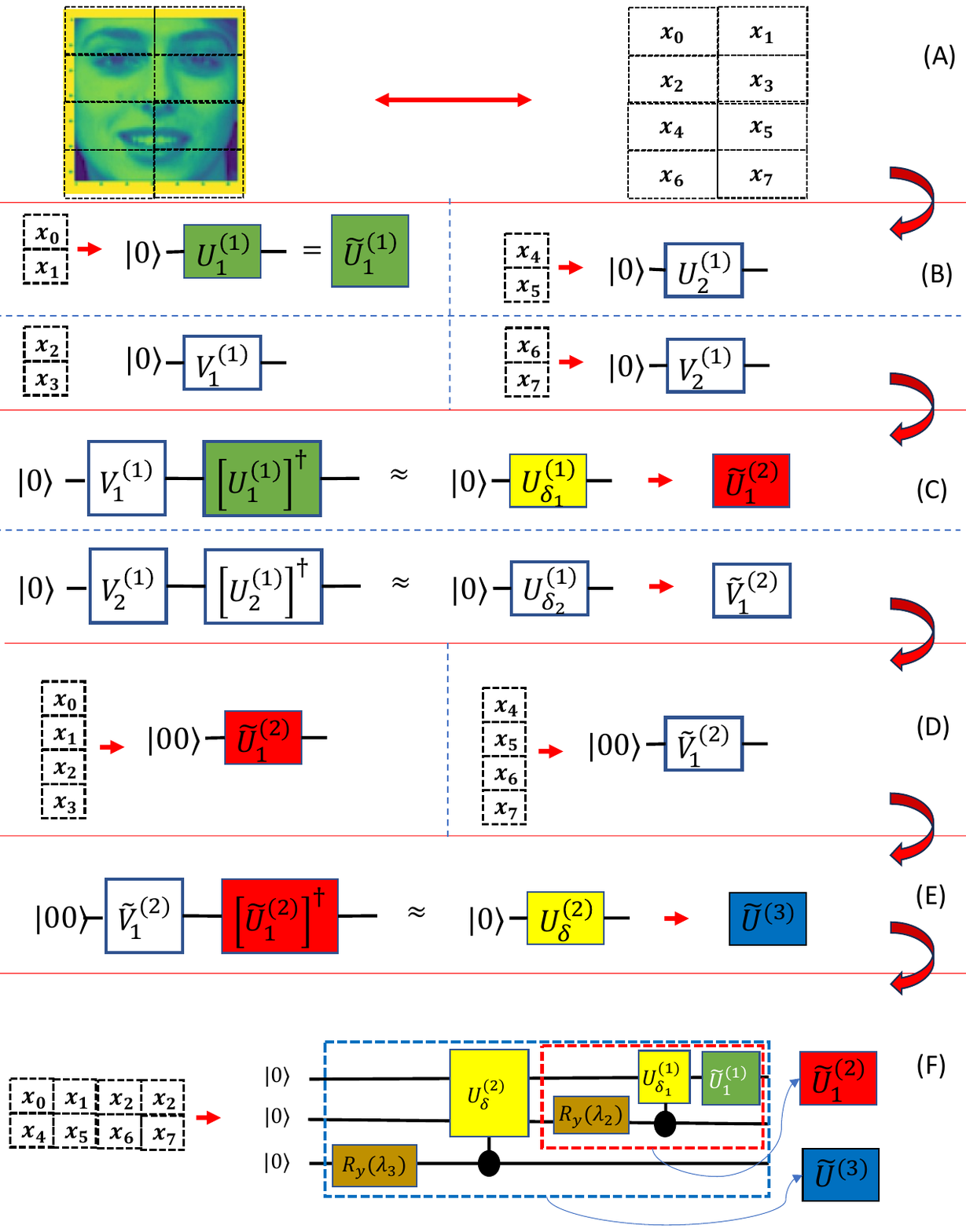}
\caption{
(A) The data is discretized into $2^n$ binary values. 
(B) Every two points are expressed in terms of a unitary operators 
$U^{(1)}_b$ and $V^{(1)}_b$ that act on a single qubit. 
All unitary operators can be found simultaneously as the segments are independent; see lines $5$ and $6$ in the algorithm of Fig. \ref{fig:main-approx-algo}. (C) Every two unitary operators $U^{(q-1)}_b$ and $V^{(q-1)}_b$ (for block $b$) from the previous step are merged, after approximation, into $U^{(q-1)}_{\delta_b}$. 
Here, the normalization of the previous step is used to compute the angles involved in the single rotation gates $R_y$ for each pair of merged unitary operators. This task corresponds to lines $11$ to  $14$ in the algorithm in Fig. \ref{fig:main-approx-algo}.
(D) The set of $\{\tilde{U}^{(q)}\}$ and $\{\tilde{V}^{(q)}\}$ are constructed from the $\{U^{(q-1)}_{\delta}\}$. The number of blocks is now reduced to half. (E) The merging and approximation continues until only one block is left. (F) At this point, all the pieces of the approximation are found and the circuit has a nested structure. The circuit depth can be constrained to stay polynomial by limiting the depth of $U_\delta$ at every step.}
\label{fig-3-2}
\end{figure*}

Here we provide an explicit example. The target data are the intensity values of a picture which we henceforth refer to as pixels. Each pixel is a nonzero real value. The pixel values form a
vector $\vec{x} = \left[x_0, \cdots, x_7\right]$, as
shown in Fig. \ref{fig-3-2}-A. The full vector is partitioned in  
$2$ blocks where each block has a size of $2^{q_{in}} \times 1 $, $q_{\rm in}=2$. Each block consists of two segments of size $2^{q_{in}-1} \times 1$. The first and second segments may be referred to as the \emph{upper} and the \emph{lower} segments, respectively. The upper and lower segments in each block $b$ are prepared by the unitary operators $U_b^{(q_{in}-1)}$ and $V_b^{(q_{in}-1)}$, respectively, as shown in Fig.  \ref{fig-3-2}-B. For example, the first block has two segments with values $(x_0, x_1)$ and  $(x_2, x_3)$ and they are encoded by the unitary operators $U_1^{(1)}$ and $V_1^{(1)}$ as
$$U_1^{(1)}\ket{0} = \frac{1}{\sqrt{x_0^2+x_1^2}} (x_0 \ket{0} + x_1 \ket{1})$$
and 
$$V_1^{(1)}\ket{0} = \frac{1}{\sqrt{x_2^2+x_3^2}} (x_2 \ket{0} + x_3 \ket{1}).$$
Since these two segments are independent, the corresponding circuits can be identified simultaneously in practice. The circuit instructions are saved (on a classical computer) as $S^{q_{\rm in}}_{U}=\left[U_1^{(1)}, V_1^{(1)}, U_2^{(1)}, V_2^{(1)}\right]$. 
Next, for each block, the corresponding $\left[U_b^{(1)}\right]^{\dagger} \,V_b^{(1)} \,\ket{0}$ is approximated and $U^{(1)}_{\delta_1}$ of block $b=1$ and $U^{(1)}_{\delta_2}$ of block $b=2$ are found. 
From the latter, the approximate circuits $\tilde{U}^{(2)}$ or $\tilde{V}^{(2)}$ are identified,  
as shown in Fig. \ref{fig-3-2}-C. See also Fig. \ref{fig-3-1} for the relation between 
$U_{\delta}$ and  $\tilde{U}$. 
Notice that the rotation angles in $R_y$ (in Fig. \ref{fig-3-1}) are computed from the normalization factors in the previous step.
For example, in $\tilde{U}^{(2)}$, the normalizations 
$N_{1u} = \sqrt{x_0^2+x_1^2}$ and $N_{1d}=\sqrt{x_2^2+x_3^2}$ are used, see Eq. (\ref{eq-x-2-lambda}).

At this stage,  
$$
\frac{1}{\sqrt{N_{1u}^2 + N_{1d}^2}}\left[x_0, \cdots, x_3\right]^T
$$
is approximated by $\tilde{U}^{(2)}\ket{00}$. Similarly, 
$$
\frac{1}{\sqrt{N_{2u}^2 + N_{2d}^2}} \left[x_4, \cdots, x_7\right]^T
$$
is approximated by $\tilde{V}^{(2)}\ket{00}$, as shown in Fig. \ref{fig-3-2}-D.

The final step is to merge $\tilde{U}^{(2)}$ or $\tilde{V}^{(2)}$. Again, the approximation approach borrowed from the MQITE algorithm to estimate $\left[U_1^{(2)}\right]^{\dagger} \,V_1^{(2)} \,\ket{00}$ is used. This results in the final $U_\delta^{(2)}$ and $\tilde{U}^{(3)}$; see Fig. \ref{fig-3-2}-E.
$\tilde{U}^{(3)}$ approximates the amplitude encoding of $\ket{x}$; the normalized values of $2^n$ pixels are thus encoded in $n$ qubits with a $\mathcal{O}(poly(n))$ circuit depth.
The final quantum circuit, which in this example is $\tilde{U}^{(3)}$, is shown in Fig. \ref{fig-3-2}-F.

\clearpage
\bibliographystyle{apsrev4-1}
\bibliography{references.bib, ref_Arslan.bib}

\end{document}